\input harvmac
\input epsf
\noblackbox


\newcount\figno
\figno=1
\def\mathrm#1{{\rm #1}}
\def\fig#1#2#3{
\par\begingroup\parindent=0pt\leftskip=1cm\rightskip=1cm\parindent=0pt
\baselineskip=11pt
\global\advance\figno by 1
\midinsert
\epsfxsize=#3
\centerline{\epsfbox{#2}}
\vskip 12pt
\centerline{{\bf Figure \the\figno} #1}\par
\endinsert\endgroup\par}
\def\bbbone {{\mathchoice {\rm 1\mskip-4mu l} {\rm 1\mskip-4mu l}
{\rm 1\mskip-4.5mu l} {\rm 1\mskip-5mu l}}}
\def\bbbc{{\mathchoice {\setbox0=\hbox{$\displaystyle\rm C$}\hbox{\hbox
to0pt{\kern0.4\wd0\vrule height0.9\ht0\hss}\box0}}
{\setbox0=\hbox{$\textstyle\rm C$}\hbox{\hbox
to0pt{\kern0.4\wd0\vrule height0.9\ht0\hss}\box0}}
{\setbox0=\hbox{$\scriptstyle\rm C$}\hbox{\hbox
to0pt{\kern0.4\wd0\vrule height0.9\ht0\hss}\box0}}
{\setbox0=\hbox{$\scriptscriptstyle\rm C$}\hbox{\hbox
to0pt{\kern0.4\wd0\vrule height0.9\ht0\hss}\box0}}}}
\def\figlabel#1{\xdef#1{\the\figno}}
\def\pano{\par\noindent}


\def\pmb#1{\setbox0=\hbox{#1}%
 \kern-.025em\copy0\kern-\wd0
 \kern.05em\copy0\kern-\wd0
 \kern-.025em\raise.0433em\box0 }
\font\cmss=cmss10
\font\cmsss=cmss10 at 7pt

\def\rlx{\relax\leavevmode}
\def\Cop{\relax\,\hbox{$\kern-.3em{\rm C}$}}
\def\Rop{\relax{\rm I\kern-.18em R}}
\def\Nop{\relax{\rm I\kern-.18em N}}
\def\Pop{\relax{\rm I\kern-.18em P}}
\def\Zop{\rlx\leavevmode\ifmmode\mathchoice{\hbox{\cmss Z\kern-.4em Z}}
 {\hbox{\cmss Z\kern-.4em Z}}{\lower.9pt\hbox{\cmsss Z\kern-.36em Z}}
 {\lower1.2pt\hbox{\cmsss Z\kern-.36em Z}}\else{\cmss Z\kern-.4em
 Z}\fi}


\def\R{{\cal{R}}}

\def\N{{\cal N}}
\def\N{{\cal N}}

\def\H{{\cal H}}

\def\A{{\cal{A}}}

\def\onth{{1\over 3}}
\def\twth{{2\over 3}}
\def\ie{{\it i.e.}}


\def\figin{\epsfcheck\figin}\def\figins{\epsfcheck\figins}
\def\epsfcheck{\ifx\epsfbox\UnDeFiNeD
\message{(NO epsf.tex, FIGURES WILL BE IGNORED)}
\gdef\figin##1{\vskip2in}\gdef\figins##1{\hskip.5in}
\else\message{(FIGURES WILL BE INCLUDED)}%
\gdef\figin##1{##1}\gdef\figins##1{##1}\fi}
\def\DefWarn#1{}
\def\figinsert{\goodbreak\midinsert}
\def\ifig#1#2#3{\DefWarn#1\xdef#1{fig.~\the\figno}
\writedef{#1\leftbracket fig.\noexpand~\the\figno}%
\figinsert\figin{\centerline{#3}}\medskip\centerline{\vbox{\baselineskip12pt
\advance\hsize by -1truein\noindent\footnotefont{\bf Fig.~\the\figno:} #2}}
\bigskip\endinsert\global\advance\figno by1}


\lref\dlm{C. Dong, H. Li, G. Mason, {\it Vertex operator algebras
associated to admissible representations of $\hat{sl}_2$},
Commun. Math. Phys. {\bf 184}, 65 (1997); {\tt q-alg/9509026}.}

\lref\dlmtwis{C. Dong, H. Li, G. Mason, {\it Twisted representations
of vertex operator algebras}, Math. Ann. {\bf 310}, 571 (1998);
{\tt q-alg/9509005}.}

\lref\gpw{A.Ch. Ganchev, V.B. Petkova, G.M.T. Watts, {\it A note on
decoupling conditions for generic level $\hat{sl}(3)_k$ and fusion
rules}, {\tt hep-th/9906139}.}

\lref\fgp{P. Furlan, A.Ch. Ganchev, V.B. Petkova, {\it $A_1^{(1)}$
admissible representations -- fusion transformations and local
correlators}, Nucl. Phys. {\bf B491}, 635 (1997); 
{\tt hep-th/9608018}.}

\lref\pry{J.L. Petersen, J. Rasmussen, M. Yu, {\it Fusion, crossing
and monodromy in conformal field theory based on $sl(2)$ current
algebra with fractional level}, Nucl. Phys. {\bf B481}, 577 (1996);
{\tt hep-th/9607129}.}

\lref\ms{G. Moore, N. Seiberg, {\it Classical and quantum conformal
field theory}, Commun. Math. Phys. {\bf 123}, 177 (1989).}

\lref\mrgone{M.R. Gaberdiel, {\it Fusion in conformal field theory as
the tensor product of the symmetry algebra},
Int. Journ. Mod. Phys. {\bf A9}, 4619 (1994); {\tt hep-th/9307183}.}

\lref\mrgtwo{M.R. Gaberdiel, {\it Fusion rules of chiral algebras},
Nucl. Phys. {\bf B417}, 130 (1994); {\tt hep-th/9309105}.}

\lref\ps{A. Pressley, G.B. Segal, {\it Loop groups}, Oxford
University Press, Oxford (1986).}

\lref\mrgwzw{M.R. Gaberdiel, {\it WZW models of general simple groups},
Nucl. Phys. {\bf B460}, 181 (1996); {\tt hep-th/9508105}.} 

\lref\mrgtwis{M.R. Gaberdiel, {\it  Fusion of twisted
representations}, Int. Journ. Mod. Phys. {\bf A12}, 5183 (1997); 
{\tt hep-th/9607036}.}

\lref\gur{V. Gurarie, {\it Logarithmic operators in conformal field
theory}, Nucl. Phys. {\bf B410}, 535 (1993); {\tt hep-th/9303160}.}

\lref\flohr{M.A. Flohr, {\it On modular invariant partition functions
of conformal field theories with logarithmic operators},
Int. J. Mod. Phys. {\bf A11}, 4147 (1996); {\tt hep-th/9509166}.}

\lref\flohrone{M. Flohr, {\it On fusion rules in logarithmic conformal
field theories}, Int. J. Mod. Phys. {\bf A12}, 1943 (1997);
{\tt hep-th/9605151}.}

\lref\RMK{M.R. Rahimi Tabar, A. Aghamohammadi, M. Khorrami, {\it The
logarithmic conformal field theories}, Nucl. Phys. {\bf B497}, 555
(1997); {\tt hep-th/9610168}.}

\lref\roh{F. Rohsiepe, {\it On reducible but indecomposable
representations of the Virasoro algebra}, {\tt hep-th/9611160}.}

\lref\flohrtwo{M. Flohr, {\it Singular vectors in logarithmic
conformal field theories}, Nucl. Phys. {\bf B514}, 523 (1998); 
{\tt hep-th/9707090}.}

\lref\gabkau{M.R. Gaberdiel, H.G. Kausch, {\it Indecomposable fusion
products}, Nucl. Phys. {\bf B477}, 293 (1996); {\tt hep-th/9604026}.}

\lref\gabkautwo{M.R. Gaberdiel, H.G. Kausch, {\it A rational
logarithmic conformal field theory}, Phys. Lett. {\bf B386}, 131
(1996); {\tt hep-th/9606050}.}

\lref\gabkauthree{M.R. Gaberdiel, H.G. Kausch, {\it A local
logarithmic conformal field theory}, Nucl. Phys. {\bf B538}, 631
(1999); {\tt hep-th/9807091}.}

\lref\ehgab{W. Eholzer, M.R. Gaberdiel, {\it Unitarity of rational
$N=2$ superconformal theories}, Commun. Math. Phys. {\bf 186}, 61
(1997); {\tt hep-th/9601163}.} 

\lref\zhu{Y. Zhu, {\it Modular invariance of characters of vertex
operator algebras}, J. Amer. Math. Soc. {\bf 9}, 237 (1996).} 

\lref\kacwac{V.G. Kac, M. Wakimoto, {\it Modular invariant
representations of infinite dimensional Lie algebras and
superalgebras}, Proc. Natl. Acad. Sci. USA {\bf 85}, 4956 (1988).}

\lref\kac{V.G. Kac, {\it Infinite-dimensional Lie algebras}, Cambridge
(1990).} 

\lref\bonn{W. Eholzer, M. Flohr, A. Honecker, R. H\"ubel, W. Nahm,
R. Varnhagen, {\it Representations of $W$-algebras with two generators 
and new rational models}, Nucl. Phys. {\bf B383}, 249 (1992).}

\lref\nahm{W. Nahm, {\it Quasi-rational fusion products},
Int. Journ. Mod. Phys. {\bf B8}, 3693 (1994); {\tt hep-th/9402039}.}

\lref\gannon{T. Gannon, {\it Modular data: the algebraic combinatorics
of conformal field theory}, {\tt math.QA/0103044}.}

\lref\ram{S. Ramgoolam, {\it New modular Hopf algebras related to
rational $k$  $\widehat{sl(2)}$}, {\tt hep-th/ 9301121}.}

\lref\berfel{D. Bernard, G. Felder, {\it Fock representations and BRST
cohomology in $sl(2)$ current algebra}, Commun. Math. Phys. 
{\bf 127}, 145 (1990).}

\lref\panda{S. Panda, {\it Fractional level current algebras and
spectrum of $C<1$ minimal models coupled to gravity}, Phys. Lett. 
{\bf B251}, 61 (1990).}

\lref\dot{V.S. Dotsenko, {\it Solving the $su(2)$ conformal field
theory with the Wakimoto free field representation}, Nucl. Phys.
{\bf B358}, 547 (1991).}

\lref\fm{B. Feigin, F. Malikov, {\it Fusion algebra at a rational level
and cohomology of nilpotent subalgebras of supersymmetric $sl(2)$}, 
Lett. Math. Phys. {\bf 31}, 315 (1994); {\tt hep-th/9310004}.}

\lref\andreev{O. Andreev, {\it Operator algebra of the $SL(2)$
conformal field theories}, Phys. Lett. {\bf B363}, 166 (1995); 
{\tt hep-th/9504082}.} 

\lref\ay{H. Awata, Y. Yamada, {\it Fusion rules for the fractional
level $\widehat{sl(2)}$ algebra}, Mod. Phys. Lett. {\bf A7}, 1185
(1992).}

\lref\malda{J. Maldacena, {\it The large N limit of superconformal
field theories and supergravity}, Adv. Theor. Math. Phys. {\bf 2}, 231
(1998); {\tt hep-th/9711200}.}

\lref\egp{J.M. Evans, M.R. Gaberdiel, M.J. Perry, {\it The No-ghost
theorem for AdS${}_3$ and the stringy exclusion principle},
Nucl. Phys. {\bf B535}, 152 (1998); {\tt hep-th/9806024}.}

\lref\deboer{J. de Boer, {\it Six-Dimensional supergravity on 
$S^3 \times AdS_3$ and 2d conformal field theory}, Nucl. Phys. 
{\bf B548}, 139 (1999); {\tt hep-th/9806104}.}

\lref\gks{A. Giveon, D. Kutasov, N. Seiberg, {\it Comments on string
theory on $AdS_3$}, Adv. Theor. Math. Phys. {\bf 2}, 733 (1998);
{\tt hep-th/9806194}.}

\lref\tesch{J. Teschner, {\it Operator product expansion and
factorization in the $H_3^+$-WZNW model}, Nucl. Phys. {\bf B571}, 555
(2000); {\tt hep-th/9906215}.}

\lref\mo{J. Maldacena, H. Ooguri, {\it Strings in $AdS_3$ and the
$SL(2,R)$  WZW model. Part 1: The spectrum}, Int. J. Mod. Phys. 
{\bf A16}, 677 (2001); {\tt hep-th/0001053}.}

\lref\mos{J. Maldacena, H. Ooguri, J. Son, {\it Strings in $AdS_3$ and
the $SL(2,R)$  WZW model. Part 2: Euclidean black hole}, 
{\tt hep-th/0005183}.}

\lref\bkog{A. Bilal, I.I. Kogan, {\it On gravitational dressing of 2D
field theories in chiral gauge}, Nucl. Phys. {\bf B449}, 569 (1995);
{\tt hep-th/9503209}.}

\lref\CKT{J.-S. Caux, I.I. Kogan, A.M. Tsvelik, {\it Logarithmic
operators and hidden continuous symmetry in critical disordered
models}, Nucl. Phys. {\bf B466}, 444 (1996); {\tt hep-th/9511134}.}

\lref\CKLT{J.-S. Caux, I.I. Kogan, A. Lewis, A.M. Tsvelik, 
{\it Logarithmic operators and dynamical extention of the symmetry
group in the bosonic $SU(2)_0$ and SUSY $SU(2)_2$ WZNW models},
Nucl. Phys. {\bf B489}, 469 (1997); {\tt hep-th/9606138}.}

\lref\kls{I.I. Kogan, A. Lewis, O.A. Soloviev, {\it
Knizhnik-Zamolodchikov-type equations for gauged WZNW models}, 
Int. J. Mod. Phys. {\bf A13}, 1345 (1998); {\tt hep-th/9703028}.}

\lref\kt{I.I. Kogan, A.M. Tsvelik, {\it Logarithmic operators in the
theory of plateau transition}, Mod. Phys. Lett. {\bf A15}, 931 (2000);
{\tt hep-th/9912143}.}

\lref\nicsan{A. Nichols, Sanjay, {\it Logarithmic operators in the
$SL(2,R)$ WZNW model}, Nucl. Phys. {\bf B597}, 633 (2001); 
{\tt hep-th/0007007}.}

\lref\lewis{A. Lewis, {\it Logarithmic CFT on the boundary and the
world-sheet}, {\tt hep-th/0009096}.}

\lref\kausch{H.G. Kausch, {\it Symplectic Fermions}, Nucl. Phys. 
{\bf B583}, 513 (2000); {\tt hep-th/0003029}.}

\lref\kawwhea{S. Kawai, J.F. Wheater, {\it Modular transformation and
boundary states in logarithmic conformal field theory},  
{\tt hep-th/0103197}.} 

\lref\kogwhea{I.I. Kogan, J.F. Wheater, {\it Boundary logarithmic
conformal field theory}, Phys. Lett. {\bf B486}, 353 (2000);
{\tt hep-th/0003184}.}

\lref\gepwit{D. Gepner, E. Witten, {\it String theory on group
manifolds}, Nucl. Phys. {\bf B278}, 493 (1986).}

\lref\kohsorba{I.G. Koh, P. Sorba, {\it Fusion rules and (sub)-modular
invariant partition functions in non-unitary theories}, Phys. Lett.
{\bf B215}, 723 (1988).}

\lref\matwal{P. Mathieu, M.A. Walton, {\it Fractional level Kac-Moody
algebras and nonunitary coset conformal field theories},
Prog. Theor. Phys. Suppl. {\bf 102}, 229 (1990).}

\lref\wzw{E. Witten, {\it Non-abelian bosonization in two dimensions}, 
Commun. Math. Phys. {\bf 92}, 455 (1984).}

\lref\walton{M.A. Walton, {\it Fusion rules in Wess-Zumino-Witten
models}, Nucl. Phys. {\bf B340}, 777 (1990).}

\lref\verlinde{E. Verlinde, {\it Fusion rules and modular
transformations in 2D conformal field theory}, Nucl. Phys. 
{\bf B300 [FS22]}, 360 (1988).}
 
\lref\gabgod{M.R. Gaberdiel, P. Goddard, {\it Axiomatic conformal
field theory}, Commun. Math. Phys. {\bf 209}, 549 (2000); 
{\tt hep-th/9810019}.}

\lref\FLM{I. Frenkel, J. Lepowsky, A. Meurman, {\it Vertex Operator
Algebras and the Monster}, Academic Press (1988).}

\Title{\vbox{
\hbox{hep--th/0105046}
\hbox{KCL-MTH-01-10}}}
{\vbox{\centerline{Fusion rules and logarithmic representations}
\vskip16pt
\centerline{of a WZW model at fractional level}  
}}
\centerline{Matthias R.\ Gaberdiel\footnote{$^\star$}{{\tt e-mail:
mrg@mth.kcl.ac.uk}}} 
\bigskip
\centerline{\it Department of Mathematics, King's College London}
\centerline{\it Strand, London WC2R 2LS, U.K.}
\smallskip
\vskip2cm
\centerline{\bf Abstract}
\bigskip
\noindent The fusion products of admissible representations of the 
$su(2)$ WZW model at the fractional level $k=-4/3$ are analysed. It is
found that some fusion products define representations for which the
spectrum of $L_0$ is not bounded from below. Furthermore, the fusion
products generate representations that are not completely reducible
and for which the action of $L_0$ is not diagonalisable. The complete
set of representations that is closed under fusion is identified, and
the corresponding fusion rules are derived. 
\bigskip

\Date{05/2001}

\newsec{Introduction}

One of the best understood conformal field theories is the WZW model
that can be defined for any (simple compact) group \refs{\wzw}. If the
so-called level is chosen to be a positive integer, the theory is
unitary and rational, and in fact these models are the paradigm for
rational conformal field theories. The fusion rules are well known
\refs{\gepwit,\kac,\walton}, and they can be obtained, via the
Verlinde formula \refs{\verlinde}, from the modular transformation
properties of the characters.  

{} From a Lagrangian point of view, the model is only well defined if
the level is integer, but the corresponding vertex operator algebra
(or the meromorphic conformal field theory in the sense of
\refs{\gabgod}) can also be constructed even if this is not the
case. Furthermore, it was realised some time ago that there exists a
preferred set of admissible (fractional) levels for which the
characters corresponding to the `admissible' representations have
simple modular properties \refs{\kacwac}. This suggests that these
admissible level WZW models define `almost' rational conformal field
theories. It is therefore interesting and important to study these
theories in order to understand to which extent results valid for
rational conformal field theories may also apply to more general
conformal field theories. 
\vskip4pt

The fusion rules of WZW models at admissible fractional level have
been studied quite extensively over the years. In particular, the
simplest case of $su(2)$ at fractional level has been analysed in
detail \refs{\berfel,\panda,\ay,\dot,\fm,\andreev,\dlm,\pry,\fgp,\gpw}
(for a good review about the various results see in particular
\refs{\gpw}). All of these fusion rule calculations essentially
determine the possible couplings of three representations. More
precisely, given two representations, the calculations determine
whether a given third representation can be contained in the fusion
product of the former two.   

Two different sets of `fusion rules' have been proposed in the
literature: the fusion rules of Bernard and Felder \refs{\berfel}
whose calculations have been reproduced in \refs{\dot,\dlm}, and the
fusion rules of Awata and Yamada \refs{\ay} whose results have been
recovered in \refs{\fm,\andreev,\pry,\fgp,\gpw}. The two calculations
differ  essentially by what class of representations is considered: in
Bernard \& Felder only admissible representations that are highest
weight with respect to the whole affine algebra (and a fixed choice of
a Borel subalgebra) are considered, while in  Awata \& Yamada also 
representations that are highest weight with respect to an arbitrary
Borel subalgebra were analysed. As a consequence, the fusion rules of
Awata \& Yamada `contain' the fusion rules of Bernard \& Felder.  

In deriving the `fusion rules' from these calculations, it is
always assumed implicitly that the actual fusion product is a direct
sum of representations of the kind that are considered. (This is to
say, there are no additional fusion channels that one has overlooked
by restricting oneself to the class of representations in question.)
In particular, in both approaches it has been assumed that the fusion
rules `close' on (conformal) highest weight representations (since
these are the only representations that were considered). However, as 
we shall explain in quite some detail, this is not true in general. In 
fact, the fusion product of two highest weight representations (with
respect to the affine algebra) contains sometimes a representation
whose $L_0$ spectrum is not bounded from
below.\footnote{$^\dagger$}{The representation has, however, the 
property that $V_n(\psi)\chi=0$ for $n\geq N$ (where $N$ depends on
both $\psi$ and $\chi$); this is sufficient to guarantee that the
corresponding correlation functions do not have essential
singularities.}  As a consequence it is not really surprising that the
fusion rules described above are somewhat incomplete.

In order to be able to analyse the fusion product without assuming
that it defines a (conformal) highest weight representation, we use the 
description of fusion that was introduced in
\refs{\ms,\mrgone,\mrgtwo,\nahm}. Refining techniques that were
developed in \refs{\nahm,\gabkau} we define a nested set of quotient
spaces of the fusion product that allows us to uncover, step by step,
more and more of the structure of the fusion product. While this
approach is necessarily incomplete (since we are not able to calculate
all such quotient spaces) it is sufficient to prove that the fusion
product is sometimes not a direct sum of (conformal) highest weight
representations. It is also sufficient to show that some of the
representations we encounter are not completely decomposable; in fact,
we shall find two indecomposable representations both of which have
the property that $L_0$ is not diagonalisable. (Representations with
this property are often called `logarithmic' representations since
their correlation functions have logarithmic branch cuts
\refs{\gur}. For some background material on this class of
representations see also
\refs{\flohr,\gabkau,\flohrone,\RMK,\roh,\flohrtwo}. For WZW models at
level $k=0$ logarithmic representations have been discovered before in 
\refs{\CKT,\CKLT}; however these models are somewhat pathological (at
$k=0$ the vacuum representation is trivial), and the relevant
logarithmic representations are quite different from what will be
analysed here.) In both cases the fact that $L_0$ is not  
diagonalisable is not visible when restricting to the highest weight
space only. (In this respect, these  representations are similar to
the logarithmic representation ${\cal R}_1$ of \refs{\gabkautwo}.)
It is therefore not surprising that these logarithmic representations
were overlooked before. On the other hand, where our calculations can
be compared with the above calculations, they reproduce the
corresponding results. 
\vskip4pt

One important insight that allows us to describe the fusion rules
fairly compactly is the observation that the fusion rules are
symmetric under the twist symmetry that originates from the outer
automorphism of the current algebra\footnote{$^\ddagger$}{At integer
level this symmetry gives rise to the so-called `simple current'
automorphism of the fusion rules.}. While we cannot prove that the
actual fusion rules respect this symmetry, we give very strong
circumstantial evidence for this claim. If this is indeed true (as we
conjecture) then it is immediate that the fusion product of certain
highest weight representations must contain representations that are
not (conformal) highest weight representations. Furthermore, this
symmetry allows us to group together all representations that are
related in this fashion. In this way we can give compact formulae for
the fusion rules (under one natural assumption that we discuss in
section~8). In particular, as was the case in \refs{\gabkautwo}, the
fusion rules close on some smaller set of representations (that
contains the two indecomposable representations together with one
irreducible representation, as well as their images under the twist
symmetry), and we find associative fusion rules for these three
representations. These fusion rules, however, bear no resemblance to
the fusion rules of either Bernard \& Felder or Awata \& Yamada, since
the three representations are quite different from those in either
\refs{\berfel} or \refs{\ay}. The corresponding $S$-matrix (that
diagonalises the fusion rules) is also different from the $S$-matrix
of Kac \& Wakimoto \refs{\kacwac}; again, this is not surprising since
the representations (and characters) are not simply the `admissible'
representations of \refs{\kacwac}.\footnote{$^\star$}{The correct
interpretation for the fusion rules that correspond to the $S$-matrix
of Kac \& Wakimoto was given in \refs{\ram}: as we shall explain in
more detail in section~8, the corresponding fusion rules agree
precisely with (a subset of) our fusion rules, and indeed support the
conjecture that the fusion rules have the aforementioned twist
symmetry.}
\vskip8pt

The paper is organised as follows. We fix our notation and describe
the twist symmetry in section~2. In section~3 we analyse which
(conformal) highest weight representations of the affine algebra are in
fact representations of the conformal field theory ({\it i.e.}
representations of the vertex operator algebra). In section~4 we
describe the algorithm for the analysis of the fusion rules in some
generality, and we apply it in section~5, 6 and 7 to the case at
hand. In section~8 we derive the full set of fusion rules (using the 
associativity of the fusion product), and section~9 contains some
conclusions. We have included two appendices where some calculations
are spelled out in some more detail.

\newsec{Notation and basic facts} 

In this paper we shall consider the WZW model corresponding to 
$su(2)$ at level $k$. The chiral algebra of this conformal field
theory contains the affine algebra $\hat{su}(2)$, whose modes satisfy
the commutation relations
\eqn\comm{\eqalign{[J^+_m,J^-_n] & = 2 J^3_{m+n} + k m \delta_{m,-n}
\cr
[J^3_m,J^\pm_n] & = \pm J^\pm_{m+n} \cr
[J^3_m,J^3_n] & = {k\over 2} m \delta_{m,-n}\,.}}
By virtue of the Sugawara construction, we can define Virasoro
generators as bilinears in the currents $J$; these Virasoro modes
satisfy the commutation relations of the Virasoro algebra
\eqn\vir{[L_m,L_n] = (m-n) L_{m+n} + {c \over 12} m (m^2-1)
\delta_{m,-n}\,,}
where $c$ is given in terms of the level $k$ as 
\eqn\central{ c= {3 k \over (k+2)}\,.}
We shall mainly consider the case $k=-4/3$ in this paper; for this
value of $k$ we have $c=-6$.

The zero modes in \comm\ satisfy the commutation relations of $su(2)$,
whose Casimir operator we denote by 
\eqn\Casimir{ C= {1\over 2} (J^+_0 J^-_0 + J^-_0 J^+_0) 
                 + J^3_0J^3_0\,.} 
Using the commutation relations \comm\ we then have 
\eqn\useful{ \eqalign{
J^+_0 J^-_0 & = C + J^3_0 - J^3_0 J^3_0 \cr
J^-_0 J^+_0 & = C - J^3_0 - J^3_0 J^3_0\,.}}
We shall often be interested in what we shall call {\it (conformal)
highest weight representations}\footnote{$^\dagger$}{Representations
with this property are often simply referred to as `highest weight
representations'. We have included the qualifier `(conformal)' in
order to distinguish these representations from the highest weight
representations of the affine algebra (which have the additional
property that the positive roots of the zero mode algebra also
annihilate $\psi$).}; these representations have the property that
they are generated by the action of the currents $J^a_m$  with $m\leq
0$ from a (conformal) highest weight state $\psi$, \ie\ a state
satisfying   
\eqn\highest{ J^a_m \psi = 0 \qquad \hbox{for $m>0$.}}
The (conformal) highest weight states in the representation generated
from $\psi$ form a representation of the zero mode algebra; if this
representation is irreducible, the Casimir operator $C$ takes a
specific value, $C_\psi$, and the conformal weight of $\psi$,
$h_\psi$, is given by    
\eqn\conf{ h_\psi = {C_\psi \over (k+2)} \,.}
\vskip4pt

In the following we shall make use of the fact that the affine algebra
has an automorphism defined by 
\eqn\auto{\eqalign{ \pi_s(J^\pm_m) & = J^\pm_{m\mp s} \cr
          \pi_s(J^3_m) & = J^3_m - {k\over 2} s\delta_{m,0}\,,}}
where $s\in\Zop$. The induced action on the Virasoro generators is
given by  
\eqn\autovir{ \pi_s(L_m) = L_m - s J^3_m 
              + {1\over 4} k s^2\delta_{m,0}\,.} 
If $s$ is even, the automorphism is inner in the sense that it can be
obtained by the adjoint action of an element in the loop group of
$SU(2)$; on the other hand, if $s$ is odd, the automorphism can be
obtained by the adjoint action of a loop in $SO(3)$ that does not
define an element in the loop group of $SU(2)$ \refs{\ps,\mrgwzw}. 

For positive integer $k$, the integrable positive energy
representations are characterised by the property that the highest
weight states transform in a representation with Casimir
$C=C(j)=j(j+1)$, where $j=0,{1\over 2}, \ldots, {k\over 2}$. In this
case, the induced action of the automorphism $\pi_1$ on the highest
weight representations is given by 
\eqn\action{\pi_1 : j \mapsto {k\over 2} - j \,.}
In particular, $\pi_s$ with $s$ even maps each integrable positive
energy representations into itself; this simply reflects the fact that
every such representation gives rise to a representation of the full
loop group, and that the automorphism for $s$ even is inner (in the
sense described above).  

Furthermore, at least for the case of positive integer $k$ where the
fusion rules are well known \refs{\gepwit}, $\pi_s$ respects the
fusion rules in the sense that  
\eqn\fus{ \left(\pi_s(\H_1) \otimes \pi_t(\H_2) \right)_{\mathrm f} 
= \pi_{s+t} \Bigl(\left( \H_1 \otimes \H_2 \right)_{\mathrm f}
\Bigr)\,.} 
This seems to be quite a general property of `twist'-symmetries such
as \auto\ (see for example \refs{\mrgtwis} for another example of this
type for the case of the ${\cal N}=2$ algebras); we shall therefore
assume in the following that the fusion rules also satisfy this
property in our case. In any case, this is consistent with what we
shall find.

\newsec{The set of allowed representations}

At $k=-4/3$, the vacuum representation has one (independent)
null-vector 
\eqn\vacnull{ \N  = \left(J^3_{-3} 
      + {3\over 2} J^+_{-2} J^-_{-1} 
      - {3\over 2} J^+_{-1} J^-_{-2}
      + {9\over 2} J^3_{-1} J^+_{-1} J^-_{-1}
      + {9\over 2} J^3_{-1} J^3_{-1} J^3_{-1}
      - {9\over 2} J^3_{-2} J^3_{-1} \right) |0\rangle \,.}
The presence of a null-vector in the vacuum representation usually
implies that only a subset of the representations of the affine
algebra actually define representations of the meromorphic conformal 
field theory. In order to determine the relevant set of 
representations, one could determine Zhu's algebra \refs{\zhu} (whose  
representations are in one-to-one correspondence with the
representations of the meromorphic conformal field
theory). In practice, however, this is quite complicated since Zhu's
algebra does not have a simple grading. Alternatively, we shall
therefore use an approach that is commonly taken in the physics
literature \refs{\bonn} (and that is believed to be equivalent to the
determination of Zhu's algebra): we shall analyse the constraint that 
comes from the condition that $V_0(\N)\psi=0$, where $\psi$ is an
arbitrary state in the representation space from which the whole
representation is generated by the action of the modes, and
$V_n(\phi)$ is the $n$-th mode of the vertex operator corresponding to
the state $\phi$ in the vacuum representation,
\eqn\vo{
V(\phi,z) = \sum_{n\in\Zop} V_n(\phi)\; z^{-n-h_\phi}\,.}
(Here $h_\phi$ is the conformal weight of $\phi$.) If the
representation is a (conformal) highest weight representation, it is  
convenient to evaluate this constraint for one of the highest weight
states. 

The complete expression for $V_0(\N)$ is quite complicated, but if we
restrict our attention to the case when $V_0(\N)$ acts on a highest
weight state, the formula simplifies quite significantly. In this case
we find (using techniques described for example in \refs{\FLM}) 
\eqn\nullrelone{\eqalign{
V_0(\N)\psi & = \left[{9\over 2} \left( J^-_0 J^+_0 J^3_0 
       + J^3_0 J^3_0 J^3_0 + J^3_0 J^3_0 \right) + J^3_0 \right] \psi \cr
     & = \left[ \left( {9\over 2} C + 1 \right) J^3_0 \right] \psi\,,}} 
where $C$ is the Casimir operator \Casimir\ and $\psi$ is an arbitrary
(conformal) highest weight state. Since we have to have that
$V_0(\N)\psi=0$ for every highest weight state in a given highest
weight representation, it follows that either the highest weight
representation is the trivial (vacuum) representation, or it has to
have Casimir equal to $-2/9$. The vacuum representation is both
highest and lowest weight with respect to the $su(2)$ zero mode
algebra, and among the representations with $C=-2/9$, there are four
representations that are either highest or lowest weight with respect
to this $su(2)$. Let us denote by $D^+_j$ the highest weight
representation that is generated from a state $|j\rangle$ satisfying
\eqn\hw{ D^+_{j}: \qquad 
\eqalign{J^+_0 |j\rangle & = 0\cr
         J^3_0 |j\rangle & = j |j\rangle \,,}}
and by $D^-_j$ the lowest weight representation that is generated from
a state $|j\rangle$ satisfying 
\eqn\hw{ D^-_{j}: \qquad 
\eqalign{J^-_0 |j\rangle & = 0\cr
         J^3_0 |j\rangle & = j |j\rangle \,.}}
The value of the Casimir for these representations is given by 
\eqn\casval{
C(D^+_j) = j (j+1) \,, \qquad C(D^-_j) = j (j-1) \,.}
Thus these representations have Casimir equal to $-2/9$ in the
following four cases
\eqn\reps{
D^+_{-\twth}\,, \qquad D^+_{-\onth} \,, \qquad
D^-_{\twth} \,, \qquad D^-_{\onth} \,.}
The two highest weight representations in \reps\ are precisely the
`admissible' representations of Kac and Wakimoto whose characters were
found to have simple modular transformation properties
\refs{\kacwac}. The fact that these admissible representations are
indeed representations of the meromorphic conformal field theory was
shown, using slightly different methods, in \refs{\dlm}. 

However, it is clear that there exist also representations with
$C=-2/9$ that are neither highest nor lowest weight with respect to
the $su(2)$ zero modes. As we shall see, one of them will play an
important role for the description of the fusion rules. This
$su(2)$ representation (which we shall denote by $E$ in the following) 
consists of the states $|m\rangle$, $m\in\Zop$, for which
\eqn\Crep{\eqalign{J^3_0 |m\rangle & = m |m\rangle\cr
             J^+_0 |m\rangle & = |m+1\rangle \cr 
             J^-_0 |m\rangle & = \left(-{2\over 9} - m (m-1)\right)
                                 |m-1\rangle \,.}}
These conditions characterise the representation $E$ uniquely. 

Obviously, these are not the only allowed representations: for every
$t\in [0,1)$ with $t\ne {1\over 3}, {2\over 3}$, there exists a
representation $E_t$ that consists of the states $|m\rangle$ with
$m-t\in\Zop$ for which the action of $su(2)$ is defined by \Crep. In
particular, this implies that the theory is not rational (as had been
previously shown in \refs{\dlmtwis}). 

On every (conformal) highest weight state, the action of $L_0$ is
proportional to the Casimir $C$. As we have seen above, in order for
the highest weight state to belong to a (non-trivial) representation
of the conformal field theory, the Casimir must take a definite value, 
$C=-2/9$. As a consequence, every allowed highest weight state is an
eigenstate of $L_0$ (with eigenvalue $h=-1/3$). Thus one may be
tempted to believe that the theory does not have any `logarithmic'
representations (which are characterised by the property that $L_0$ is
not diagonalisable). Quite surprisingly, this is however not true. As
we shall see, the fusion product of two highest weight representations
contains a representation that is {\it not} a (conformal) highest
weight representation, and for which the action of $L_0$ is not
diagonalisable.  
\vskip4pt

All of these highest weight representations have non-trivial null
vectors. At grade one, the null vector is explicitly given by 
\eqn\nullgradeone{
\N_1 =  \left[ \left( 9 m^2 - 1\right) J^3_{-1} 
+ {9\over 4} \left(2m+{2\over 3}\right) J^+_{-1} J^-_0
+ {9\over 4} \left(2m-{2\over 3}\right) J^-_{-1} J^+_0 \right] 
|m\rangle\,,}
where $|m\rangle$ has $J^3_0$ eigenvalue $m$ (and Casimir
$C=-2/9$). In fact, $\N_1 = V_{-1}(\N) |m\rangle$, where $\N$ is the 
vacuum null vector \vacnull. There is also a non-trivial null vector
at grade two (\ie\ a null-vector that is not a descendant of $\N_1$);
it is given by $\N_2=V_{-2}(\N)|m\rangle$, and its explicit expression
is 
\eqn\nullgradetwo{\eqalign{
\N_2  = & \left[ \left( 9 m^2 + {9\over 2} m - 1 \right) J^3_{-2} 
+ {3\over 2} (3m+2) J^+_{-2} J^-_0 
+ {3\over 2} (3m-2) J^-_{-2} J^+_0 \right. \cr
& \quad \left.
+ {27\over 2} m J^3_{-1} J^3_{-1} 
+ {9\over 2} m J^-_{-1} J^+_{-1}  
+ {9\over 2} J^3_{-1} J^+_{-1} J^-_0 
+ {9\over 2} J^3_{-1} J^-_{-1} J^+_0 \right] |m\rangle \,.}}
\vskip4pt

Some of the representations \reps\ are interrelated by the
automorphisms $\pi_s$. In order to describe these relations, let us
introduce the following notation. If $\rho:{\cal A}\rightarrow
\hbox{End}(\H)$ describes the representation $\H$, then we denote by
$\pi_s(\H)$ the representation that is defined by
$\rho\circ\pi_s$. With this notation we then have 
\eqn\autorels{\eqalign{ \pi_1(\H_0) & = D^-_{\twth} \cr
                        \pi_{-1}(\H_0) & = D^+_{-\twth} \cr
                        \pi_1(D^+_{-\onth}) & = D^-_{\onth}\,.}}
Here $\H_0$ denotes the vacuum representation. The first two lines
imply that $\pi_2(D^+_{-\twth})=D^-_{\twth}$. Apart from these special
cases, the application of an automorphism to any of these 
representations typically leads to a representation that does not have
the (conformal) highest weight property. However, all representations 
that arise in this fashion have the property that for a given state
$\psi$ in the representation and a given $\chi$ in the vacuum
representation, there exists a positive integer $N$ (that depends on 
$\psi$ and $\chi$) such that 
\eqn\vanishing{ V_n(\chi) \psi = 0 \qquad \hbox{for all $n>N$.}}
This truncation property is sufficient to guarantee that all
correlation functions will only have poles (rather than essential
singularities).

\newsec{The fusion algorithm}

Before we begin to describe our results in detail, let us briefly
explain some of the techniques that we shall be using in the
following. As was explained in \refs{\mrgone,\mrgtwo,\nahm} fusion can
be defined in terms of a ring-like tensor product: given two
representations of the chiral algebra, $\H_1$ and $\H_2$, the tensor
product $(\H_1\otimes\H_2)$ carries two natural actions of the chiral
algebra (that are defined by the comultiplication formulae
\refs{\ms,\mrgone,\mrgtwo,\nahm}), and the fusion product
$(\H_1\otimes\H_2)_{\mathrm f}$ is the quotient space of the tensor
product where we identify these two actions. The main advantage of this
description relative to most other approaches to fusion is that we do
not presuppose that the fusion product has any specific properties. In
most other fusion calculations one only analyses whether one of the
familiar (highest weight) representations is contained in the fusion
product, but here we shall analyse the fusion product itself, not just
some subrepresentations it may contain. In fact, while in all cases
that have been analysed so far, the fusion product of two highest
weight representations is highest weight, there is no abstract reason
why this has to be so, and indeed, in the present context we shall find
that this is not the case. As we shall see, the fusion product of two
(conformal) highest weight representations defines a representation
that is not a (conformal) highest weight representation. We shall 
also find that the fusion product of certain (conformal) highest
weight representations defines a reducible but indecomposable
representation for which the action of $L_0$ is not diagonalisable.

In principle, one would like to describe the full space
$(\H_1\otimes\H_2)_{\mathrm f}$ directly, but unfortunately, this is
a fairly hopeless task. Instead, we shall therefore analyse a number
of nested quotient spaces; these will give more and more information
about the fusion product and will allow us to show certain properties
(and to make very strongly supported conjectures for others). The
most important quotient space is the space that we obtain by
quotienting out all states that can be obtained by the action of the
negative modes ${\cal A}_{-}$ (\ie\ the modes $V_n(\phi)$ with $n<0$)
\refs{\nahm},  
\eqn\highest{ \left( \H_1 \otimes \H_2\right)^{(0)}_{\mathrm f} 
\equiv \left( \H_1 \otimes \H_2\right)_{\mathrm f} / 
{\cal A}_{-} \left( \H_1 \otimes \H_2\right)_{\mathrm f} \,,}
where the action of the chiral algebra on the fusion product is
defined in terms of the comultiplication formulae
\refs{\ms,\mrgone,\mrgtwo}. It is clear that this space carries a
natural action of the zero modes. Furthermore, the space is naturally
dual to the highest weight space of the conjugate representation, and 
therefore, for an irreducible highest weight representation, can be
identified with the highest weight space itself; in the following we
shall therefore sometimes refer to it as the `highest weight space of
the fusion product'. It can be efficiently computed using the
algorithm described in \refs{\gabkau} (see also \refs{\nahm}):
choosing suitable insertion points for the two representations
($z_1=1, z_2=0$), we have the identities   
\eqn\simple{ \eqalign{(\bbbone\otimes J^a_{-n}) & \cong 
- (J^a_0 \otimes \bbbone) + ({\cal A}_+ \otimes \bbbone) \cr
(J^a_{-n} \otimes \bbbone) & \cong 
- (-1)^{n} (\bbbone\otimes J^a_0 ) + (\bbbone\otimes {\cal A}_+) 
\,,}}
where ${\cal A}_+$ is the subalgebra of positive modes and we have
assumed that $n\geq 1$. Thus if we are evaluating \highest\ on 
(conformal) highest weight states in $\H_1$ and $\H_2$, the second
terms in \simple\ vanish.

As an aside, one can use this approach to analyse the allowed
representations of the conformal field theory, following an idea
described in \refs{\ehgab}. To this end, we consider the fusion
product of an arbitrary (conformal) highest weight representation with
the vacuum representation, and analyse the conditions under which the
original highest weight representation is contained in this fusion
product, and therefore in \highest. Since the vacuum representation
has a null vector \vacnull, we obtain a non-trivial constraint from 
using \simple\ repeatedly,
\eqn\calcone{\eqalign{0 & = (\psi \otimes \N) \cr
& \cong \left(-J^3_0 + {3\over 2} J^-_0 J^+_0 
- {3 \over 2} J^-_0 J^+_0 - {9\over 2} J^-_0 J^+_0 J^3_0
- {9\over 2} J^3_0 J^3_0 J^3_0 - {9\over 2} J^3_0 J^3_0
\right) \psi \otimes |0\rangle \cr
& \cong - \left(\left[\left(1+{9\over 2} C\right) J^3_0 \right]\psi 
\otimes |0\rangle \right) \,,}}
where $C$ is the Casimir operator \Casimir\ and $\psi$ is an arbitrary
highest weight vector. This reproduces \nullrelone.
\vskip4pt

As we shall see in the next section, the knowledge of \highest\ is
sometimes not sufficient to characterise the fusion product
uniquely. When appropriate we shall therefore also consider a slightly
larger quotient space (that therefore captures slightly more information
about the fusion product). In particular, we shall consider the
quotient space 
\eqn\nextorder{
 \left( \H_1 \otimes \H_2\right)^{(+1)}_{\mathrm f} 
\equiv \left( \H_1 \otimes \H_2\right)_{\mathrm f} / 
{\cal A}^{+1}_{-} \left( \H_1 \otimes \H_2\right)_{\mathrm f} \,,}
where ${\cal A}^{+1}_{-}$ is the algebra that is spanned by the modes 
\eqn\Aplusone{
{\cal A}^{+1}_{-}: \qquad \qquad \eqalign{ 
J^+_{-n} \qquad & \hbox{with $n\geq 2$} \cr
J^3_{-n} \qquad & \hbox{with $n\geq 1$} \cr
J^-_{-n} \qquad & \hbox{with $n\geq 1$}\,.}}
It is easy to see that ${\cal A}^{+1}_{-}$ closes among these
modes. It is also clear that \highest\ is a natural quotient space of
\nextorder; in this sense the two quotient spaces are nested. Finally,
it is worth mentioning that $L_0$, $J^3_0$ and $J^-_0$ (but not
$J^+_0$) act in a well-defined manner on this quotient space. 

In order to determine \nextorder\ we can use essentially the same
algorithm as before for \highest. In fact, \simple\ still holds
provided that $n\geq 1$ for $a=3,-$ and $n\geq 2$ for $a=+$. In
addition we have the identity 
\eqn\lesssimple{
(J^+_{-1}\otimes \bbbone) \cong (\bbbone\otimes J^+_{-1}) + 
\sum_{m=0}^{\infty} (-1)^m (J^+_{m}\otimes \bbbone) + 
\sum_{l=0}^{\infty}  (\bbbone\otimes J^+_{l})\,.}
The last identity comes from the fact that, on the fusion product, 
$\Delta_{1,0}(J^+_{-1})$ and $\widetilde{\Delta}_{0,-1}(J^+_{-1})$
only differ by states that lie in the quotient space of
\nextorder. (Here we have used the notation of
\refs{\mrgone,\mrgtwo}.) 

Ideally one would like to determine yet bigger quotient 
spaces. However, the complexity of the calculation increases very 
quickly, and the above is essentially the limit of what can be
calculated feasibly by hand. An implementation of the calculation on a
computer is not straightforward since all highest weight spaces are
infinite dimensional, and thus the computer algorithm used in
\refs{\gabkau} cannot be applied directly. At any rate, the above
quotient spaces are already sufficient to show that the fusion product
of certain highest weight representations contains a logarithmic
representation.

\newsec{The fusion rules}

We are now in the position to work out the various fusion products. 
Let us begin by determining the `highest weight space' \highest\ of
the fusion product of $D^+_{-\twth}$ with an arbitrary (conformal)
highest weight representation $\H$. First of all, we can use \simple\
repeatedly to reduce any state in the fusion product (up to states in
the quotient space) to a sum of products of highest weight states.
Next we want to obtain the constraints that follow from the existence
of the null vectors in $D^+_{-\twth}$. To this end, it is useful to 
observe that for $m=-\twth$, 
\eqn\nulltwth{
\N^+_{-\twth}= J^+_0 \N_1 = - J^+_{-1} \left| -\twth\right\rangle }
is also a null-vector in $D^+_{-\twth}$ (as one can easily check
directly). Using \simple\ together with this null-vector we get the
condition   
\eqn\calctwo{\eqalign{0 & = (\psi\otimes \N^+_{-\twth}) \cr
& \cong \left( J^+_0 \psi \otimes \left|-\twth\right\rangle\right)\,,}}
where $\psi\in\H^{(0)}$ is an arbitrary (conformal) highest weight
state in $\H$. (Here and in the following we shall denote by
$\H^{(0)}$ the subspace of (conformal) highest weight states of $\H$.)
If $\H=D^+_{-\onth}$ or $\H=D^+_{-\twth}$, then every
$\phi\in\H^{(0)}$ can be written as $\phi=J^+_0 \psi$ for some
$\psi\in\H^{(0)}$. Thus we find that for every $\phi\in\H^{(0)}$,
$(\phi\otimes |-2/3\rangle)$ lies in the quotient by which we divide
to obtain the highest weight subspace of the fusion
product. Furthermore, using recursively the relation  
\eqn\recur{
0 \cong J^-_0 \left(\phi\otimes\left|-\twth\right\rangle\right)
= \left(J^-_0 \phi\otimes \left|-\twth\right\rangle\right)
+ \left(\phi\otimes J^-_0 \left|-\twth\right\rangle\right)\,,}
we can show that all states in the tensor product of the highest
weight space of $D^+_{-\twth}$ and $\H^{(0)}$ are in the quotient
space by which we divide.  Thus we conclude that the fusion product of
$D^+_{-\twth}$ with $D^+_{-\twth}$ or $D^+_{-\onth}$ does not contain
{\it any} (conformal) highest weight states,   
\eqn\resultone{ 
\left(D^+_{-\twth} \otimes D^+_{-\twth} \right)^{(0)}_{\mathrm f} = 
\left(D^+_{-\twth} \otimes D^+_{-\onth} \right)^{(0)}_{\mathrm f} = 0
\,.} 
A similar conclusion was also reached in \refs{\ram}. This is in fact
in agreement with the symmetry \fus\ since
$D^+_{-\twth}=\pi_{-1}(\H_0)$, and we therefore expect that  
\eqn\resultonep{\eqalign{
\left(D^+_{-\twth} \otimes D^+_{-\twth} \right)_{\mathrm f} & =
\pi_{-1}\left( D^+_{-\twth} \right) \cr
\left(D^+_{-\twth} \otimes D^+_{-\onth} \right)_{\mathrm f} & =
\pi_{-1}\left( D^+_{-\onth} \right)\,.}}
Indeed, it is easy to see that the right hand side of \resultonep\
does not define a (conformal) highest weight representation.

We can test \resultonep\ further by determining the quotient space 
\nextorder\ that captures more than just the highest weight states. 
For the case when both representations are $D^+_{-\twth}$ we have done
this calculation, and we have found that\footnote{$^\ddagger$}{Strictly
speaking, our calculation only allows us to derive an upper bound on
the size of the quotient space. However, we have used all available
null vector relations, and we are therefore confident that this bound
is actually saturated. This comment applies equally to all other
calculations of quotient spaces in this paper.}  
\eqn\resulconf{
\left(D^+_{-\twth} \otimes D^+_{-\twth} \right)^{(+1)}_{\mathrm f} =  
\left\{ \left(\hat{m},\hat{m}\right): \; \hat{m} = -{4\over 3}, 
- {7\over 3}, \ldots \right\}\,,}
where the first entry of $\left(\hat{m},\hat{m}\right)$ refers to its
$J^3_0$ eigenvalue, and the second to its $L_0$ eigenvalue, 
\eqn\eigenvals{ \eqalign{
J^3_0 \left(\hat{m},\hat{m}\right) & = \hat{m}
\left(\hat{m},\hat{m}\right) \cr
L_0 \left(\hat{m},\hat{m}\right) & = \hat{m} 
\left(\hat{m},\hat{m}\right)\,.}}
This agrees with the spectrum of $J^3_0$ and $L_0$ on the quotient
space of $\pi_{-1}( D^+_{-\twth})$ where we divide out the image of
${\cal A}_{-}^{+1}$: the states that survive in this quotient space
are the image (under $\pi_{-1}$) of the original highest weight states
in $D^+_{-\twth}$. Using \auto\ and \autovir\ the spectrum of these
states is then precisely described by \resulconf\ and \eigenvals.
\vskip4pt

If $\H=D^-_{\twth}$, then the first part of the argument is similar,
except now there is one state in the highest weight space of
$D^-_{\twth}$ that cannot be written as $J^+_0\phi$: this is the state
$|2/3\rangle$. Using \recur\ as before this allows us then to show
that all tensor products of highest weight states for which the total
$J^3_0$ eigenvalue is bigger than zero lie in the quotient by which we
divide. (Here the total $J^3_0$ eigenvalue is the sum of the two
$J^3_0$ eigenvalues.) We can similarly use the null-vector $\N_1$ 
of $D^-_{\twth}$ (or rather, as before in \nulltwth, the null vector
that is obtained from $\N_1$ by the action of $J^-_0$) to deduce that
the same holds for those tensor products whose total $J^3_0$
eigenvalue is less than zero. This leaves us with the states in the 
tensor product for which the total $J^3_0$ eigenvalue is zero: these
are the states of the form 
\eqn\states{(J^+_0)^l \left|\twth\right\rangle \otimes
(J^-_0)^l \left| -\twth\right\rangle\,,}
where $l=0,1,\ldots$. By considering 
\eqn\calcthree{
0 = \left(J^-_0\right)^{m+1} 
\left(\left(J^+_0\right)^m \left|\twth\right\rangle \otimes 
J^+_{-1}\left| -\twth\right\rangle\right)}
and using the algorithm above, we obtain a recursion relation for the
state with $l=l_0$ in \states\ in terms of states with
$l=0,1,\ldots,l_0-1$. Thus the highest weight space is actually
one-dimensional, and it consists of a state of $J^3_0$ eigenvalue
zero. This suggests that we have 
\eqn\resulttwo{
\left(D^+_{-\twth} \otimes D^-_{\twth} \right)_{\mathrm f}= \H_0 \,.}
We have also checked that the null-vectors $\N_2$ for both
representations do not give rise to any additional constraints. 
\vskip4pt

\noindent Finally we find, using similar arguments,
\eqn\resultthree{\eqalign{
\left(D^+_{-\twth} \otimes D^-_{\onth} \right)_{\mathrm f} 
& =  D^+_{-\onth}\cr
\left(D^-_{\twth} \otimes D^+_{-\onth} \right)_{\mathrm f} 
& =  D^-_{\onth}\,.}}
All of these results are in agreement with \fus\ and \autorels, and in
fact could have been deduced from this symmetry. 

The only fusion products involving the representations \reps\ that are 
not determined in terms of this symmetry are the fusion products
involving $D^+_{-\onth}$ and $D^-_{\onth}$. Using the algorithm
described above we have calculated the highest weight space 
\eqn\zwischen{
\left(D^+_{-\onth} \otimes D^-_{\onth}\right)^{(0)}_{\mathrm f}
= \{ |0\rangle\} \oplus E \,,}
where $|0\rangle$ is a state with $J^3_0$ and $L_0$ eigenvalue zero,
and $E$ is the representation that was discussed in
section~3. (This result agrees with the $x=\infty$ limit (that
corresponds to a specific choice for the Borel subalgebra for one of
the three representations) described in \refs{\gpw}: the
representation $E$ does not appear in their fusion product since $E$
is not an (affine) highest weight representation.) Furthermore, we
have checked that   
\eqn\zwischenn{
\left(D^+_{-\onth} \otimes D^-_{\onth}\right)^{(+1)}_{\mathrm f}
= \Bigl\{ \left(m,m\right): m=0,1,2,\ldots\Bigr\} \bigoplus E \,,} 
where both the $J^3_0$ and $L_0$ eigenvalue of
$(m,m)$ is $m$. Taking these results together this suggests that the
actual fusion product is    
\eqn\resultfour{
\left(D^+_{-\onth} \otimes D^-_{\onth}\right)_{\mathrm f}
= \H_0 \oplus \H_E \,,}
where $\H_E$ is the representation of the affine algebra whose
(conformal) highest weight space $\H_E^{(0)}$ is $E$. Indeed, since 
$\H_E$ has a null vector at grade one, $\N_1$, for which the
coefficient of $J^+_{-1}$ does not vanish (see \nullgradeone), only
the ground states of $\H_E$ contribute in \zwischenn; on the other
hand, the first term in \zwischenn\ corresponds to the states of the
form  $(J^+_{-1})^l|0\rangle$ that survive the quotient by the image
of  $\A_{-}^{+1}$ in the vacuum representation. Using the
automorphism symmetry \fus\ this result then also implies  
\eqn\resultfive{\eqalign{
\left(D^+_{-\onth} \otimes D^+_{-\onth}\right)_{\mathrm f}
& = D^+_{-\twth} \oplus \pi_{-1}(\H_E) \cr
\left(D^-_{\onth} \otimes D^-_{\onth}\right)_{\mathrm f}
& = D^-_{\twth} \oplus \pi_{1}(\H_E) \,.}}
Both of these results are again in agreement with the direct
calculation of their highest weight spaces, 
\eqn\zwischenzwei{\eqalign{
\left(D^+_{-\onth} \otimes D^+_{-\onth}\right)^{(0)}_{\mathrm f}
& = \left(D^+_{-\twth}\right)^{(0)}\cr
\left(D^-_{\onth} \otimes D^-_{\onth}\right)^{(0)}_{\mathrm f}
& = \left(D^-_{\twth}\right)^{(0)}\,,}}
since the highest weight space of $\pi_{\pm 1}(\H_E)$ is empty. Again,
these calculations are in agreement with the $x=0$ limit of the
calculation described in \refs{\gpw}. 

\newsec{The fusion of $\H_E$ with $D^{\pm}_{\mp\onth}$}

So far we have described all fusion products involving the original
representations in \reps. We have found that the fusion rules do not
close among these representations (and their images under $\pi_s$)
alone, but rather that we generate the representation $\H_E$ (as
well as its images under $\pi_s$). In order to describe the full
fusion ring we therefore need to study the fusion of $\H_E$ with all
representations in \reps\ as well as itself. First we consider the
fusion of $\H_E$ with $D^{\pm}_{\mp\twth}$ for which the symmetry
\fus\ predicts 
\eqn\resultseven{
\left(D^{\pm}_{\mp\twth} \otimes \H_E\right)_{\mathrm f} =
\pi_{\mp 1} (\H_E) \,.}
Again, we have checked that the highest weight spaces agree on both
sides (both are empty). This leaves us with the fusion of $\H_E$ with
$D^{\pm}_{\mp\onth}$ which we denote as 
\eqn\resultseven{\eqalign{
\left(D^{+}_{-\onth} \otimes \H_E\right)_{\mathrm f} & = 
\R_{-\onth} \cr
\left(D^{-}_{\onth} \otimes \H_E\right)_{\mathrm f} & = 
\R_{\onth} = \pi_1\left(\R_{-\onth}\right)\,.}}
In relating the two results we have again used the symmetry \fus.

The analysis of $\R_{-\onth}$ is actually quite subtle, and we
therefore describe it in some detail. First we determine the `highest 
weight space' of $\R_{-\onth}$, \ie\ the quotient of $\R_{-\onth}$ by
the states that lie in the image of $\A_{-}$ using the algorithm
described above. We find that
\eqn\zwischenzwei{
\left(\R_{-\onth}\right)^{(0)} = \left\{ 
\left(m-\onth,-\onth\right) : m\in\Zop
\right\}\,,}
where the action of the zero modes on $(m-\onth,-\onth)$ is defined
by  
\eqn\zeromodes{\eqalign{
J^3_0 \left(m-\onth,-\onth\right) & = \left(m-\onth\right) 
\left(m-\onth,-\onth\right)  \cr
L_0 \left(m-\onth,-\onth\right) & = - \onth 
\left(m-\onth,-\onth\right) \cr
J^+_0 \left(m-\onth,-\onth\right) & =  
\left(m+1-\onth,-\onth\right) \cr
J^-_0 \left(m-\onth,-\onth\right) & = -(m-1)\left(m-\twth\right) 
\left(m-1-\onth,-\onth\right)\,.}}
This representation of $su(2)$ is reducible, but not decomposable:
because of the factor of $(m-1)$ on the right hand side of the last
equation, the states $(m-\onth,-\onth)$ with $m\geq 1$ form an
irreducible subrepresentation (that is equivalent to
$(D^-_{\twth})^{(0)}$), but the complementary space does not define a
representation since $J^+_0(-\onth,-\onth)=(\twth,-\onth)$. The
structure of this representation can schematically be described by  
\ifig\Ceinsdrei{The structure of $C_{-\onth}=(\R_{-\onth})^{(0)}$. The   
arrows describe the action of
$J^\pm_0$.}{\epsfxsize4.0in\epsfbox{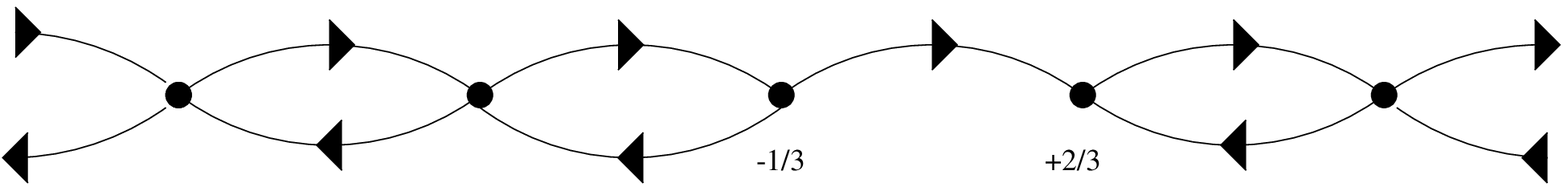}}       

We can similarly determine the `highest weight space' of $\R_{\onth}$ 
and we find that 
\eqn\zwischendrei{
\left(\R_{\onth}\right)^{(0)} = \left\{ 
\left(\onth + m,-\onth\right) : m\in\Zop
\right\}\,,}
where now 
\eqn\zeromodesprime{\eqalign{
J^3_0 \left(m+\onth,-\onth\right) & = \left(m+\onth\right) 
\left(m+\onth,-\onth\right)  \cr
L_0   \left(m+\onth,-\onth\right) & = - \onth 
\left(m+\onth,-\onth\right) \cr
J^-_0 \left(m+\onth,-\onth\right) & =  \left(m-1+\onth,-\onth\right) \cr
J^+_0 \left(m+\onth,-\onth\right) & = -(m+1)\left(m+\twth\right) 
\left(m+1+\onth,-\onth\right)\,.}}
Again the states $(m+\onth,-\onth)$ with $m\leq -1$ form an
irreducible subrepresentation (that is equivalent to
$(D^+_{-\twth})^{(0)}$), but the complementary space does 
not define a representation since
$J^-_0(\onth,-\onth)=(-\twth,-\onth)$.  Its structure is described by
the `mirror' image of that sketched in Figure~1.

Based on these results one could believe that the representations
$\R_{\mp \onth}$ are simply the affine representations whose actual
highest weight spaces are the indecomposable $su(2)$ representations
that are described by  \zeromodes\ and \zeromodesprime,
respectively. However, this is {\it not} correct. One of the reasons
why one may be suspicious about this conjecture is that the two
representations $\R_{\mp \onth}$ would then {\it not} be related by
$\pi_{\pm 1}$, and thus the symmetry \fus\ would not hold any more for
\resultseven. In fact, if one postulates that $\R_{\mp \onth}$ 
{\it are} related by $\pi_{\pm 1}$, it is easy to see that neither of
the two representations can be a highest weight representation. 

In order to analyse the situation further we have therefore determined
the quotient space \nextorder\ of the first fusion product in
\resultseven, and we have found that 
\eqn\nextresult{\eqalign{
\left( \R_{-\onth} \right)^{(+1)} & = 
\left\{ \left(m-\onth,-\onth\right) : m\in\Zop, m\ne 0\right\}
\bigoplus
\left\{ \left(m-\onth,m-\onth\right) : m\in\Zop, m\ne 0\right\} \cr
& \qquad \bigoplus
\left\{ \left(-\onth,-\onth\right)_1, \left(-\onth,-\onth\right)_2, 
\left(-\onth,\twth\right)\right\} \,.}}
As before, the labels of the different states characterise their
$J^3_0$ and $L_0$ eigenvalues, except for the two states
$(-\onth,-\onth)_1$ and  $(-\onth,-\onth)_2$ for which the action of
$L_0$ is given by
\eqn\central{ \eqalign{
L_0 \left(-\onth,-\onth\right)_1 & = 
- \onth \left(-\onth,-\onth\right)_1 + \left(-\onth,-\onth\right)_2
\cr 
L_0 \left(-\onth,-\onth\right)_2 & = 
- \onth \left(-\onth,-\onth\right)_2\,.}}
In particular, this implies that the representation $\R_{-\onth}$ is a
logarithmic representation. Since this is one of the central results of
this paper, we shall describe its derivation in some more detail in
the appendix. 

Roughly speaking, the space $(\R_{-\onth})^{(+1)}$ consists of the
states in $(\R_{-\onth})^{(0)}$ -- these are the states
$\left(m-\onth,-\onth\right)$ -- as well as the states in 
$\pi_{-1}((\R_{\onth})^{(0)})$ -- these are the states 
$\left(m-\onth,m-\onth\right)$. The latter space is naturally a
quotient space of $(\R_{-\onth})^{(+1)}$ since under $\pi_{-1}$, the
subspace that is generated by the negative modes becomes the space
that is generated by $J^+_{-n}$ with $n\geq 2$, $J^3_{-n}$ with 
$n\geq 1$ and $J^-_{-n}$ with $n\geq 0$. The relation to 
$(\R_{\onth})^{(0)}$ (under the map $\pi_{-1}$) suggests 
that $J^+_{-1}$ and $J^-_{1}$ are the `step'-operators that move 
between the states $\left(m-\onth,m-\onth\right)$. Because of the
structure of \zeromodesprime\ we then know that 
\eqn\observe{
J^+_{-1}\left(-{4\over 3},-{4\over 3} \right)\cong 0}
in the quotient space of $(\R_{-\onth})^{(+1)}$ by the action of
$J^-_0$. (This is the quotient space that corresponds to 
$\pi_{-1}((\R_{\onth})^{(0)})$.) On the other hand, we can calculate
the action of $J^-_0$ on $(\R_{-\onth})^{(+1)}$, and we find (for some
suitable normalisation) $J^-_0 (m-\onth,-\onth) = (m-1-\onth,-\onth)$
provided that $m\ne 1,0$. Furthermore, if $m=1$ we have 
$J^-_0 (\twth,-\onth) = (-\onth,-\onth)_2$. This is in agreement with
the description of \zeromodes\ since, as is shown in the appendix, 
$(-\onth,-\onth)_2\cong 0$ in the `highest weight space'. This
suggests that we must have (up to some constant that we can absorb
into the definition of the states)
\eqn\guessone{
J^-_0 \left(\twth,-\onth\right) = 
J^+_{-1} \left(-{4\over 3},-{4\over 3} \right)\,.}
Finally, we know from the analysis of the highest weight space that 
$J^+_0 \left(-\onth,-\onth\right)_1=\left(\twth,-\onth\right)$, and
from the analysis of $\pi_{-1}((\R_{\onth})^{(0)})$ that 
$J^-_{1} \left(-\onth,-\onth\right)_1=
\left(-{4\over 3},-{4\over 3}\right)$. Thus \guessone\ implies
\eqn\guesstwo{
J^-_0 J^+_0 \left(-\onth,-\onth\right)_1 = 
{1\over 3} \left(-\onth,-\onth\right)_2 = 
J^+_{-1} J^-_{1} \left(-\onth,-\onth\right)_1 \,.}
where we have made a specific prediction for the relative
normalisation constants (that will be justified further below). 
The resulting structure is summarised in Figure~2. 

\ifig\Rdrei{The structure of the generating states of the
representation ${\cal R}_{-\onth}$. Here we have arranged the
circles representing the states according to their charges, with the
horizontal axis corresponding to $J^3_0$, and the vertical axis to
$L_0$. The horizontal array of circles represent the states of the
form $(m-\onth,-\onth)$ with $m\in\Zop$, while the diagonal array 
of circles represents the states $(m-\onth,m-\onth)$, $m\in\Zop$. The
two arrays intersect at $(-\onth,-\onth)$, and the circle at this
intersection corresponds to the state $(-\onth,-\onth)_1$. The arrows
indicate the action of $J^\pm_0$ (for the horizontal line) and
$J^\pm_{\mp 1}$ (for the diagonal line).  Finally, the empty circle
represents the state $(-\onth,-\onth)_2$ whose position in the charge
lattice has been slightly shifted so that it does not lie on top of
the other state with these
charges.}{\epsfxsize3.5in\epsfbox{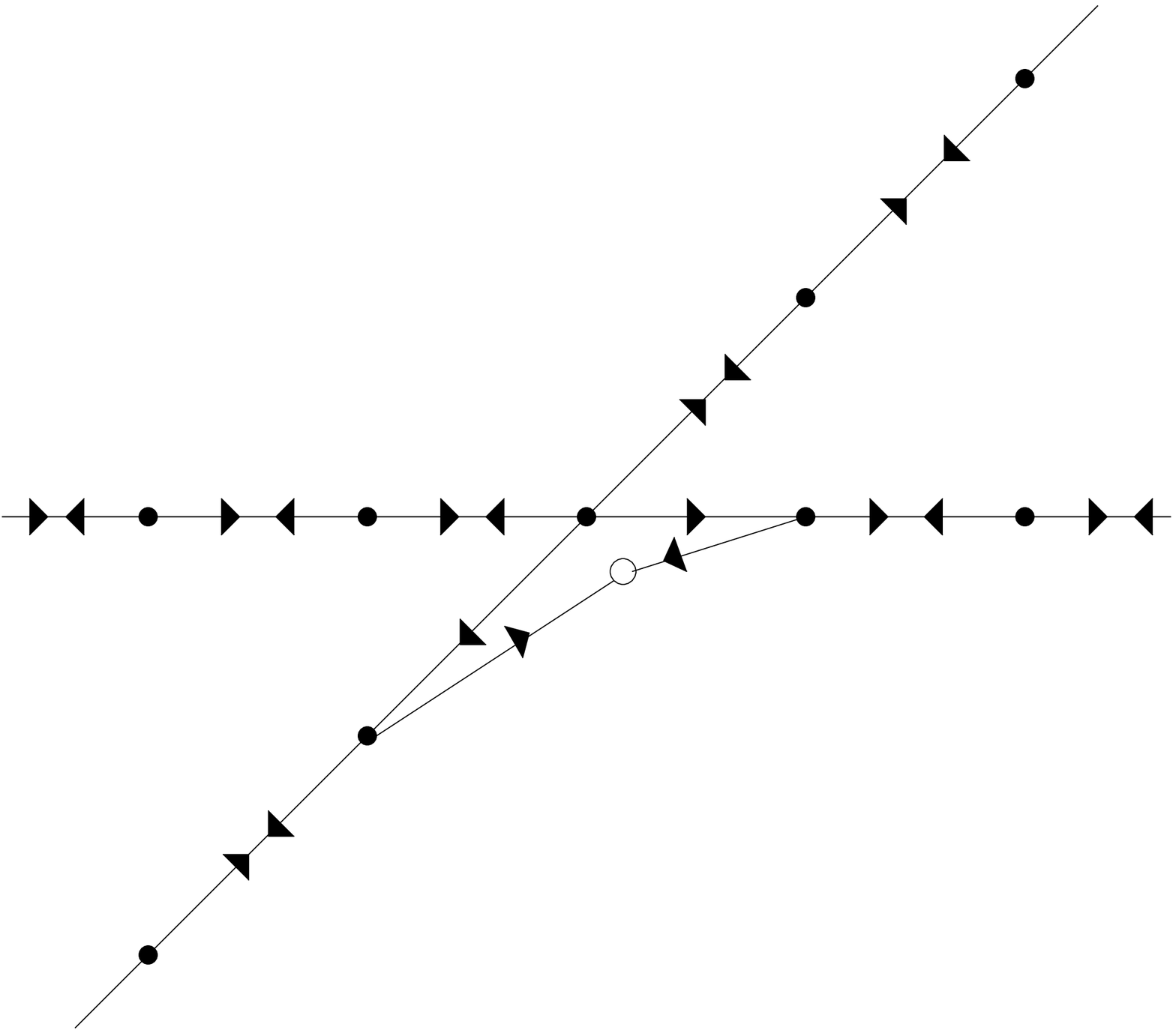}}   
\vskip4pt

On the basis of what we have determined we cannot expect to be able to
derive the structure of $\R_{-\onth}$ completely; however, we can make
an ansatz for its structure, and check it against the various pieces
of evidence that we have accumulated. In making this ansatz we shall
be guided by the principle that the representation $\R_{-\onth}$ is as
well behaved as it can possibly be. For example, we know that
$\left(-\onth,-\onth\right)_1$ is not a (conformal) highest weight
state (since it is not annihilated by $J^-_{1}$), but we can make the
ansatz that it is as close to being (conformal) highest weight as
possible by postulating  
\eqn\postulate{ \eqalign{
J^+_{n} \left(-\onth,-\onth\right)_1 = 0  \qquad & 
\hbox{for $n\geq 1$}  \cr
J^3_{n} \left(-\onth,-\onth\right)_1 = 0  \qquad & 
\hbox{for $n\geq 1$}  \cr
J^-_{n} \left(-\onth,-\onth\right)_1 = 0  \qquad & 
\hbox{for $n\geq 2$} \cr
J^a_n J^+_0 \left(-\onth,-\onth\right)_1 = 0  \qquad & 
\hbox{for $n\geq 1$.}}}

As a first piece of evidence in favour of this ansatz we want to 
show that the resulting representation is actually an allowed
representation of the conformal field theory. To this end, we
want to check that $V_0(\N) \left(-\onth,-\onth\right)_1=0$ (where
$\N$ is again the vacuum null vector), but now taking into account
that $\left(-\onth,-\onth\right)_1$ is not a highest weight. Because
of this modification we now get (instead of \nullrelone)
\eqn\consist{\eqalign{
0 = & \Bigl[{9\over 2} \left( J^-_0 J^+_0 J^3_0 
       + J^3_0 J^3_0 J^3_0 + J^3_0 J^3_0 \right) + J^3_0 \cr
    & \qquad 
       + {9\over 2} J^3_{-1} J^-_1 J^+_0 
       + {9\over 2} J^+_{-1} J^-_{1} J^3_0 
       + 3 J^+_{-1} J^-_{1}\Bigr] \left(-\onth,-\onth\right)_1 \cr
  = & {3\over 2} \left[J^+_{-1} J^-_1 - J^-_0 J^+_0 \right] 
 \left(-\onth,-\onth\right)_1\,.}}
Thus \consist\ follows from \guesstwo. We can similarly check that  
$V_1(\N) \left(-\onth,-\onth\right)_1=0$, but this only reproduces the
same constraint. Also, $V_L(\N) \left(-\onth,-\onth\right)_1=0$ for
$L\geq 2$ follows automatically from our ansatz \postulate.

Given that the representation satisfies this consistency condition, we
can then construct a null-vector in $\R_{-\onth}$ by applying
$V_{-1}(\N)$ to $\left(-\onth,-\onth\right)_1$, say (and taking care,
again, of the fact that this vector only satisfies \postulate). The
null vector we obtain in this way is explicitly given as 
\eqn\nullnew{
\widehat{\N} = \left[ -3 J^-_{-1} J^+_0 + 3 J^+_{-2} J^-_{1}
+ {9\over 2} J^3_{-1} J^+_{-1} J^-_1\right] 
\left(-\onth,-\onth\right)_1 \,.}
By applying $J^-_{1}$ to $\widehat{\N}$ we then get a non-trivial
relation in $\left(\R_{-\onth}\right)^{(+1)}$, namely
\eqn\relin{\eqalign{
0 & = \left[ -3 J^-_1 J^-_{-1} J^+_0 + 3 J^-_1 J^+_{-2} J^-_{1}
+ {9\over 2} J^-_1 J^3_{-1} J^+_{-1} J^-_1\right] 
\left(-\onth,-\onth\right)_1 \cr
& \cong {9 \over 2} J^-_0 J^+_{-1} J^-_1 \left(-\onth,-\onth\right)_1
\cr 
& = {9 \over 2} J^-_0 \left(-\onth,-\onth\right)_2\,.}}
This relation is crucial for explaining why the space of states with
quantum numbers $(-{4\over 3},-\onth)$ is one-dimensional in 
$\left(\R_{-\onth}\right)^{(+1)}$ (rather than two-dimensional as one
may have naively thought). 

It is also worth mentioning that the null vector $\widehat{\N}$ does
not involve the state $J^+_{-1} J^-_0 \left(-\onth,-\onth\right)_1$;
this is presumably the reason why this vector is {\it not} removed
from $(\R_{-\onth})^{(+1)}$. (This is the state with quantum numbers  
$(-\onth,\twth)$.) 

The analysis for $\R_{\onth}$ is completely analogous, and its
structure is described by the `mirror' image of Figure~2.

\newsec{The fusion of $\H_E$ with itself}

Finally, we need to analyse the fusion of $\H_E$ with itself. The
highest weight space of this fusion product agrees precisely with the
right hand side of \zwischen, and one may therefore expect that the
fusion of $\H_E$ with itself is precisely $\H_0\oplus\H_E$. However,
given that the fusion of $\H_E$ with $D^+_{-\onth}$ contains an
indecomposable representation, one may expect that also the fusion of
$\H_E$ with itself may be indecomposable. In order to test this
further we have  determined the quotient space \nextorder, and we have
found   
\eqn\zwischennn{
\left(\H_E \otimes \H_E\right)^{(+1)}_{\mathrm f}
= \Bigl\{ \left(m,m\right): m\in\Zop\ldots\Bigr\} \bigoplus E \,,} 
where both the $J^3_0$ and $L_0$ eigenvalue of
$(m,m)$ is $m$. This differs crucially from \zwischenn\ in that now
$m$ runs over all integers (rather than just the non-negative
integers). Furthermore, we have determined the action of $J^+_{-1}$ on
these states --- this is well-defined on the quotient space
\zwischennn\ ---  and we have found that, for some suitable
normalisation,  
\eqn\actionn{
J^+_{-1} (m,m) = {1\over 3} (m+1)(3m+1) (m+1,m+1)\,.}
In particular, it therefore follows that the state $(0,0)$ is not of
the form $J^+_{-1}(-1,-1)$, in agreement with the result for the
highest weight space. By symmetry, it also follows that the analogous
quotient space where we exchange $J^+$ and $J^-$ has the same
structure; in particular, we have states $(-m,m)$ for which
\eqn\actionnn{
J^-_{-1} (-m,m) = {1\over 3} (-m+1)(-3m+1) (-m-1,m+1)\,.}
Summarising what we have found so far we therefore propose that 
\eqn\HE{
\left(\H_E \otimes \H_E\right)_{\mathrm f} = \R_0 \oplus \H_E\,,} 
where $\R_0$ is an extension of the vacuum representation.

The structure of the states $(-m,m)$ is the same as that of 
$\pi_{-1}(C_{\twth})$, where $C_{\twth}$ is the other indecomposable
representation of the zero mode algebra whose underlying vector space
is the same as $C_{-\onth}$; its structure is schematically described 
by 
\ifig\Czweidrei{The structure of $C_{\twth}$. The arrows describe the
action of $J^\pm_0$.}{\epsfxsize4.0in\epsfbox{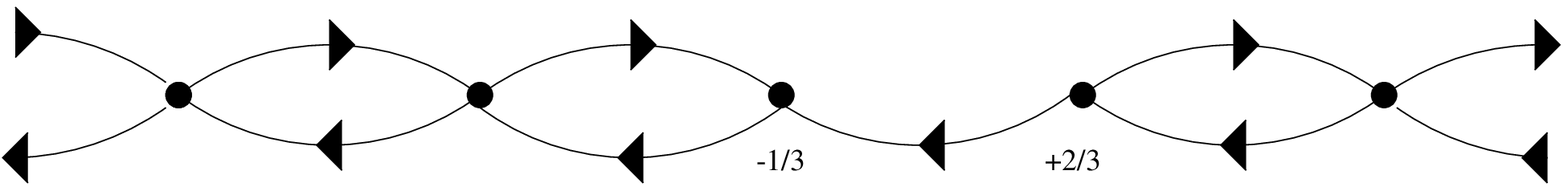}} 
Similarly, the states of the form $(m,m)$ have the same structure as 
$\pi_1(C_{-\twth})$, where $C_{-\twth}$ is the `mirror image' of 
$C_{\twth}$. Roughly speaking, the representation $\R_0$ is therefore
the combination of these two representations. More precisely, we claim
that $\R_0$ is generated by the action of the affine modes from
$\omega$, where 
\eqn\structureone{\eqalign{
J^+_{1} \omega & = (1,-1) \cr
J^-_{1} \omega & = (-1,-1)\,,}}
and
\eqn\structuretwo{\eqalign{
J^-_{-1} J^+_{1} \omega & = \Omega = J^+_{-1} J^-_{1} \omega \cr
J^-_0 J^+_0 \omega & = \gamma \Omega = J^-_0 J^+_0 \omega \cr
L_0 \omega & = \left(3+{3\over 2} \gamma\right)\Omega\,,}}
where $\gamma\ne 0,-2$ is some constant.\footnote{$^\star$}{The
representations that correspond to different values for $\gamma$ are
inequivalent; the actual representation that occurs in the fusion
product therefore has a specific value of $\gamma$. Unfortunately,
this constant cannot be determined from the knowledge of the various
quotient spaces that we have calculated.}  The structure of this 
representation is schematically described by Figure~4.
\ifig\Rnull{The structure of the generating states of the
representation ${\cal R}_{0}$. As before, the different states
have been arranged according to their $J^3_0$ and $L_0$ charge. The
two diagonal arrays of points represent the representations
$\pi_{-1}(C_{\twth})$ and  $\pi_1(C_{-\twth})$, while the horizontal
line corresponds to the states that can be obtained by $J^\pm_0$ from
$\omega$. The empty circle represents the state $\Omega$ whose
position in the charge  lattice has been slightly shifted so that it
does not lie on top of the state
$\omega$.}{\epsfxsize4.5in\epsfbox{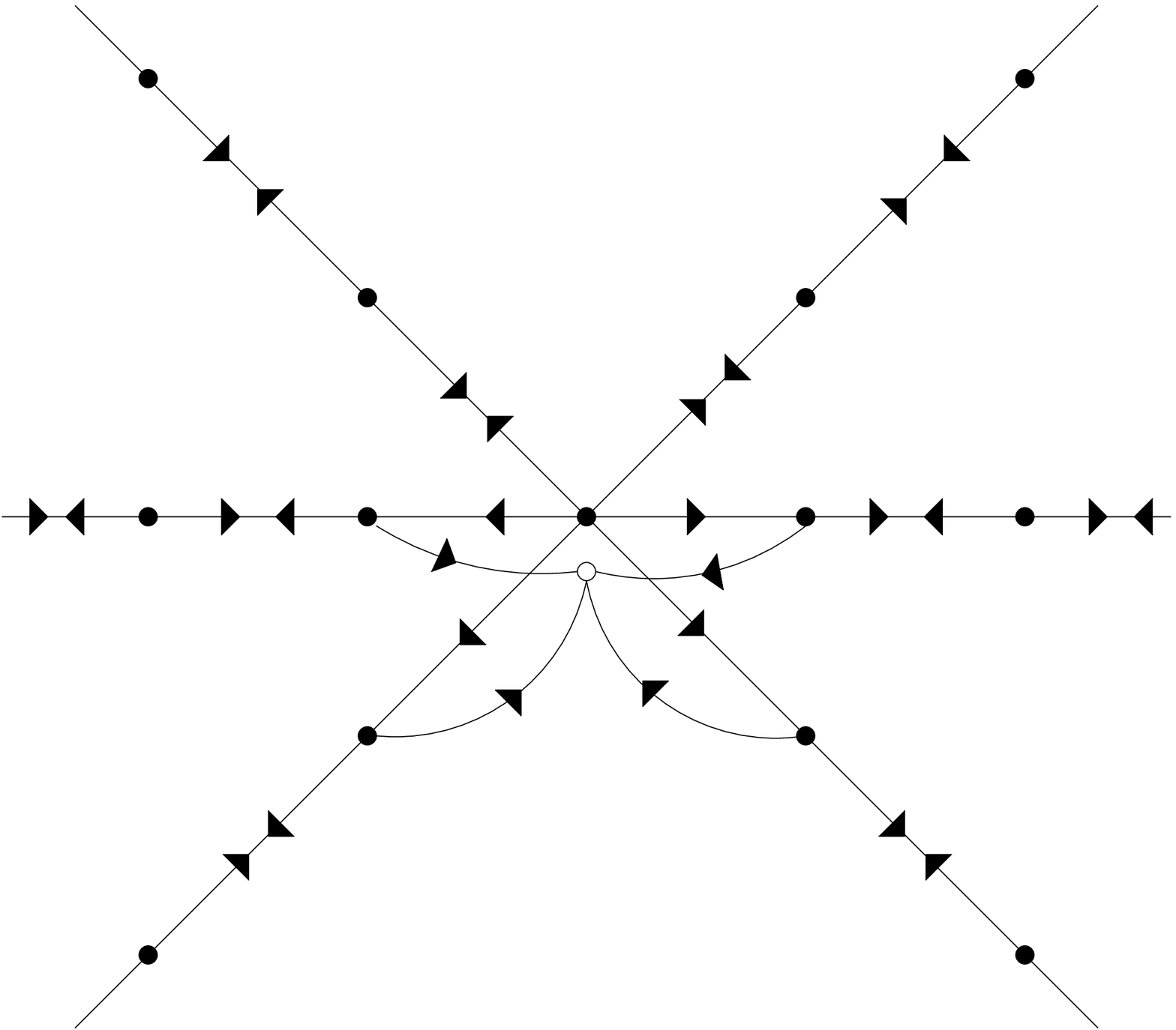}}  
The fact that $J^+_0 J^-_0 \omega = J^-_0 J^+_0 \omega$ follows from
the commutation relations of the affine algebra together with
$J^3_0\omega=0$. If we postulate that $\omega$ is annihilated by  
\eqn\postulatetwo{\eqalign{
J^a_n \omega & = 0 \qquad n\geq 2 \cr
J^3_1 \omega & = 0 \cr
J^+_1 J^-_1 \omega & = 0 \cr
J^\mp_1 J^\pm_0 \omega & = 0 \,,}}
then the condition $J^-_{-1}J^+_{1}\omega=J^+_{-1}J^-_{1}\omega$ is
again a consequence of $V_1(\N)\omega=0$. Since $\Omega$ is in the
image of both $J^+_{-1}$ and $J^-_{-1}$, it does not appear in the
quotient space \nextorder, in agreement with the result \zwischenn. We
have therefore calculated the quotient space where we divide out the
image of $J^3_{-n}$, $J^\pm_{-n-1}$ with $n\geq 1$, and we have found,
that at $J^3_0$ charge zero, the relevant quotient space of the fusion
product of $\H_E$ with itself is six-dimensional: three of these six
states have conformal weight $-\onth$, $\twth$ and ${5\over 3}$, and
correspond to states in $\H_E$. Of the remaining three states, one has
conformal weight $2$, while the other two form a Jordan block of
length two at $h=0$ --- these correspond to the states $\omega$ and
$\Omega$. (In particular, this implies that $\gamma\ne -2$.) We have
also checked that if we divide out further by the image of $J^a_0$,
then the Jordan block disappears; this implies that $\Omega$ must be
in the image of $J^\pm_0$ (and therefore that $\gamma\ne 0$).

\newsec{Fusion closure}

Up to now we have described all fusion products involving the original 
representations in \reps\ as well as $\H_E$. However, as we have shown
in the previous sections, the fusion rules still do not close on this
set of representations since we generate the representations
$\R_{-\onth}$ and $\R_0$ (as well as their images under
$\pi_s$). Given that the structure of $\R_{-\onth}$ and $\R_0$ is
quite complicated it would be very difficult to establish directly the
fusion products involving these representations. However, we can use
the fact that the fusion product is associative \refs{\mrgone} to
predict some other fusion products; furthermore, {\it all} fusion
products can be determined once we have made a (fairly natural)
assumption about one additional fusion product.

First of all it follows from the associativity of the fusion product
that  
\eqn\assone{\eqalign{
\left( D^-_{\onth} \otimes \R_{-\onth} \right)_{\mathrm f} & = 
\left[\left( D^-_{\onth} \otimes D^+_{-\onth} \right)_{\mathrm f}
\otimes \H_E \right]_{\mathrm f} \cr
& = \left[\left( \H_0 \oplus \H_E \right)_{\mathrm f}
\otimes \H_E \right]_{\mathrm f} \cr
& = 2 \, \H_E \oplus \R_0 \,.}}
In order to determine the remaining fusion products we now make the
conjecture that 
\eqn\asstwo{
\left( D^+_{-\onth} \otimes \R_0 \right)_{\mathrm f}  = \R_{-\onth}
\,.} 
This is a very natural assumption since $\R_0$ is an extension of the
vacuum representation, and one would therefore expect that the
right-hand side of \asstwo\ is also an extension of $D^+_{-\onth}$;
of the representations we have considered so far, the representation
$\R_{-\onth}$ is the only representation with this property. We can
also give some direct evidence for this conjecture by calculating the
highest weight space of the fusion product in \asstwo; this is done in
appendix~B.

With this assumption we can then determine the remaining fusion rules,
using the associativity of the fusion product\footnote{$^\dagger$}{As 
we shall see, the resulting `fusion rules' are all non-negative
integers; this is by no means guaranteed by our procedure, and
therefore provides a consistency check on our assumption \asstwo. In
fact, if we had assumed that the right-hand-side of \asstwo\ was just 
$D^+_{-\onth}$ the resulting `fusion rules' would lead to some negative
integers.}. For example, we have
\eqn\assthree{
\left( \R_{-\onth} \otimes \H_E \right)_{\mathrm f} = 
\left[D^+_{-\onth} \otimes \left(\H_E \otimes\H_E \right)_{\mathrm f} 
\right]_{\mathrm f} = 2\, \R_{-\onth}\,.}
We also find that 
\eqn\assfour{\eqalign{
\left[ \H_E \otimes 
\left( D^-_{\onth} \otimes \R_{-\onth}\right)_{\mathrm f} 
\right]_{\mathrm f} & = 
2 \left( \H_E \otimes \H_E \right)_{\mathrm f} \bigoplus 
\left( \H_E \otimes \R_0\right)_{\mathrm f} \cr
& = 2 \, \H_E \oplus 2\, \R_0 \bigoplus 
\left( \H_E \otimes \R_0\right)_{\mathrm f}\,,}}
while on the other hand we have
\eqn\assfive{\eqalign{
\left[ \H_E \otimes 
\left( \R_{-\onth} \otimes D^-_{\onth} \right)_{\mathrm f} 
\right]_{\mathrm f} & = 
\left[ \left( \H_E \otimes \R_{-\onth} \right)_{\mathrm f} 
\otimes D^-_{\onth} \right]_{\mathrm f} \cr
& = 2 \, \left(  \R_{-\onth} \otimes D^-_{\onth}\right)_{\mathrm f}
\cr 
& = 4 \, \H_E \oplus 2 \R_0 \,,}}
from which we can conclude that 
\eqn\asssix{
\left(\H_E \otimes \R_0\right)_{\mathrm f} = 2\, \H_E \,.}
Using similar techniques we also find that 
\eqn\assseven{\eqalign{
\left( \R_0 \otimes \R_{-\onth} \right)_{\mathrm f} & = 2\, 
\R_{-\onth} \cr
\Bigl( \R_0 \otimes \R_0 \Bigr)_{\mathrm f} & = 2\, \R_0 \cr
\left( \R_{\onth} \otimes \R_{-\onth} \right)_{\mathrm f} & = 2\, 
\R_0 \oplus 4 \H_E\,.}}

Similar to what happened in \refs{\gabkautwo} we observe that the
fusion rules close among the representations $\R_0$, $\R_{\onth}$ and
$\H_E$, together with their images under $\pi_s$. (In particular, this
set also includes the representation
$\R_{-\onth}=\pi_{-1}(\R_{\onth})$.) In some sense this is again the 
natural set of representations to consider since $\R_0$ (and its
images under $\pi_{\mp 1}$) contains $\H_0$ (and $D^\pm_{\mp\twth}$)
as a subrepresentation, and $\R_{\pm\onth}$ contains
$D^\mp_{\pm\onth}$ as subrepresentations. Furthermore, these 
representations (together with their images under $\pi_{\pm 1}$)
contain all the different indecomposable representations of the zero
mode algebra for which the Casimir takes value $C=-2/9$ and $J^3_0$ is
in $\Zop/3$. Indeed there exist four such representations, namely
$C_{\pm\onth}$ and $C_{\pm\twth}$; the former two are contained in
$\R_{\pm\onth}$, while the latter two arise in $\pi_{\pm 1}(\R_0)$. It
is therefore quite natural that the fusion rules should close on this
set of representations.

Given that the fusion rules observe the symmetry \fus, we can group
the representations into orbits under $\pi_s$\footnote{$^\ddagger$}{We
do not propose that these orbits form representations of some loop
group. We simply choose to combine these inequivalent representations
(of the affine algebra) in order to obtain a compact formula for the
fusion rules. The actual fusion rules (including the appropriate
action of $\pi_s$) are described by the various formulae above.}. Thus
we have three orbits whose fusion closes among itself; the relevant
fusion rules can then compactly be described by   
\eqn\fusionresults{\eqalign{
[\R_0] \otimes [\R_0] &= 2\, [\R_0] \cr
[\R_0] \otimes [\H_E] &= 2\, [\H_E] \cr
[\R_0] \otimes [\R_{\onth}] &= 2\, [\R_{\onth}] \cr
[\H_E] \otimes [\H_E] &= [\H_E] \oplus [\R_0] \cr
[\H_E] \otimes [\R_{\onth}] &= 2\, [\R_{\onth}] \cr
[\R_{\onth}] \otimes [\R_{\onth}] &= 2\, [\R_0] \oplus 4\, [\H_E]\,.}} 
It is not difficult to check that the resulting fusion rules are
associative and commutative; the $S$-matrix that diagonalises these
fusion rules is then given by 
\eqn\Smatrix{
S= \pmatrix{ {1\over 2 \sqrt{2}} & {1\over 2 \sqrt{2}} & 
{1\over \sqrt{2}} \cr
{1\over \sqrt{2}} & {1\over \sqrt{2}} & -{1\over \sqrt{2}} \cr
{\sqrt{3}\over 2 \sqrt{2}} & -{\sqrt{3}\over 2 \sqrt{2}} & 0}\,.}
This $S$-matrix is not unitary let alone symmetric, and the
fusion rules are therefore not `self-dual' in the sense of Gannon
\refs{\gannon}. Given that the $S$-matrix of Kac \& Wakimoto
\refs{\kacwac} is in general ({\it i.e.} for higher algebras) not
unitary \refs{\matwal}, the absence of unitary may not be 
surprising. The lack of symmetry may be related to the fact that 
the (unspecialised) character of the representation $\H_E$ vanishes
for $0<z<1$. Unfortunately, since we do not know the characters of
$\R_0$ and $\R_{\onth}$, we cannot check whether this $S$ matrix has
an interpretation in terms of the modular transformation of these
characters. Also, it is not clear whether it is appropriate to group
these representations into the above orbits. For example, there are
arguments why it may be natural to consider the orbits of $\pi_{2s}$
or  $\pi_{3s}$ (where $s$ is again an integer) instead; on the other
hand, this does not seem to improve the structural properties of the
$S$-matrix: in either case the resulting $S$-matrix is not unitary
nor symmetric.

Obviously, this $S$-matrix is not at all similar to the $S$-matrix
obtained by Kac \& Wakimoto \refs{\kacwac}. However, this is not
surprising since the latter has an interpretation in terms of fusion
rules that correspond to a {\it different} subset of representations
(that also closes under fusion). As was pointed out by Ramgoolam
\refs{\ram}, we have the (formal) character identity 
\eqn\ramone{
\chi_{D^+_{-\onth}} (\tau,z) = - \chi_{D^-_{\twth}} (\tau,z)\,.}
Thus, up to some trivial signs, the $S$-matrix of Kac \& Wakimoto
equally describes the modular transformation rules of the three
representations  $\H_0$, $D^\pm_{\mp\twth}$. Once these signs are
appropriately introduced, the fusion rules that are associated to this
modified $S$-matrix via the Verlinde formula \refs{\verlinde} define
non-negative integer fusion coefficients\footnote{$^\star$}{The fusion
rule coefficients that are obtained from the unmodified Kac-Wakimoto 
$S$-matrix are not non-negative integers as was first observed by
\refs{\kohsorba} (see also \refs{\matwal}).}  
\eqn\fusionrulesram{\eqalign{
[D^+_{-\twth}] \otimes [D^-_{\twth}] & = [\H_0] \cr
[D^+_{-\twth}] \otimes [D^+_{-\twth}] & = [D^-_{\twth}] \cr
[D^-_{\twth}] \otimes [D^-_{\twth}] & = [D^+_{-\twth}]\,.}}
From the point of view of the modular properties it is natural to
group together representations that are related by $\pi_{3s}$ since
only then the conformal weight of the corresponding representations is 
guaranteed to differ by an integer (see \auto\ with
$k=-4/3$). Actually more is true in the present case since it follows
from the explicit formula for the characters \refs{\kacwac} that 
\eqn\characters{
\chi_{\pi_s(\H_0)} (\tau;z) = \left\{ 
\eqalign{ \chi_{\H_0} (\tau;z) \qquad & \hbox{if $s=0$ mod $3$} \cr
\chi_{D^-_{\twth}} (\tau;z) \qquad & \hbox{if $s=1$ mod $3$} \cr
\chi_{D^+_{-\twth}} (\tau;z) \qquad & \hbox{if $s=-1$ mod $3$.}}
\right. }
Thus the equivalence classes in \fusionrulesram\ correspond naturally
to the images of the representations under $\pi_{3s}$. These fusion
rules then agree precisely with what we have claimed above; in fact,
they are a direct consequence of \fus\ together with \autorels.

\newsec{Conclusions}

In this paper we have determined the fusion rules of $su(2)$ at the
fractional level $k=-4/3$. Starting with the `admissible' 
highest weight representations we have shown that we generate
representations that are not (conformal) highest weight representations.
This is an immediate consequence of the automorphism symmetry \fus\
which we have confirmed in a number of cases directly. We have also
found that some of the fusion products are not completely
decomposable, and that they contain representations for which the
action of $L_0$ is not diagonalisable. We have found a set of three
representations (two of which are indecomposable) whose fusion closes
among itself (together with its images under the action of the
automorphism). 

There are a few obvious directions in which the results of this paper
should be extended. First of all, it would be important to understand
the structure of the various indecomposable representations that we
have found in more detail. This will presumably require a calculation
of some even larger quotient spaces (that uncover more of the
structure of the resulting representation). All of the calculations
that we have done in this paper were in essence done by hand; in order
to make further progress it is presumably necessary to implement these
calculations on a computer. 

It would also be interesting to understand the fusion of $su(2)$ for
the other admissible fractional levels ($k=-4/3$ is only the simplest
such example).  Furthermore, it would be interesting to understand
what happens for algebras of larger rank, such as $su(3)$, {\it etc}. 

At any rate, the results of this paper suggest that the fusion of the
admissible representations of all fractional level WZW  models will
contain indecomposable (and more specifically logarithmic)
representations. This seems to indicate that, despite what one may 
have thought originally, `logarithmic representations' do occur quite
generically. 
\vskip4pt

One of the original motivations for this work was the technical
similarity between the representation theory for fractional level
$su(2)$ and for the WZW model that corresponds to the non-compact
group $SU(1,1)$. The latter model is believed to describe string
compactification on $AdS_3$, and therefore plays an important role in 
the analysis of the AdS/CFT correspondence 
\refs{\malda,\egp,\deboer,\gks,\tesch,\mo,\mos}. There are some
indications that the fusion rules of the WZW model based on the group
$SU(1,1)$ do indeed contain logarithmic representations  
\refs{\bkog,\kls,\kt,\nicsan,\lewis}. It would be interesting to see
whether the techniques used above can shed further light on this
issue. It would also be interesting to study D-branes in these
backgrounds. For the case of the local logarithmic theory constructed 
in \refs{\gabkauthree} (see also \refs{\kausch}) the construction of
boundary states has recently been accomplished in
\refs{\kawwhea}, following the programme outlined in
\refs{\kogwhea}. It would be interesting to see whether an analogous
construction can be performed for the fractional level WZW models, or
indeed $SU(1,1)$.

\vskip 1cm

\centerline{{\bf Acknowledgements}}\pano

\noindent I thank Alexander Ganchev, Terry Gannon, Andreas Recknagel,
Mark Walton and G\'erard Watts for useful discussions, and in
particular Alexander Ganchev and Terry Gannon for much needed
encouragement.  

\noindent I am grateful to the Royal Society for a University Research    
Fellowship. 

\vskip 1cm

\appendix{A}{The fusion calculation for $\R_{-\onth}$}

In this appendix we want to give some of the details of the derivation
of our central result \nextresult\ and \central. First we want to
describe the space 
\eqn\appone{
\left(D^{+}_{-\onth} \otimes \H_E\right)^{(+1)}_{\mathrm f}\,.}
Because of the null vector $\N_1$, we can rewrite any state of the
form $(J^+_{-1})^l |m\rangle$ in terms of states that do not involve
any $J^+_{-1}$ modes; thus we can apply the relations in \simple\ to
show that \appone\ is spanned by states in the tensor product of the
two highest weight spaces. Next we want to derive further restrictions 
by using the null vectors of the two representations. It is most
convenient to consider a null-vector (at grade two) that does not
involve any $J^+_{-1}$ modes (since these cannot easily be `removed'
using the algorithm). By taking a suitable combination of descendants
of $\N_1$ and $\N_2$ we find that the representation $D^+_{-\onth}$
has the null vector
\eqn\apptwo{\eqalign{
\N_{-\onth}  =  & \left(
9 J^3_{-1} J^3_{-1} J^-_0 J^-_0 - 45 J^3_{-2} J^-_0 J^-_0 
+ 14 J^-_{-1} J^-_{-1} + 36 J^-_{-1} J^3_{-1} J^-_0 \right. \cr
& \left.  \qquad 
- 48 J^-_{-2} J^-_0 + 9 J^+_{-2} J^-_0 J^-_0 J^-_0 \right)
\left|-\onth\right\rangle \,.}}
Using the algorithm \simple\ we then obtain (after some algebra) the
relation 
\eqn\appthree{\eqalign{
0 = & \Bigl[ 
-{14 \over 81} (3j+2) (3j+1) (3j-1) (3j-2) (\bbbone\otimes\bbbone) 
+ {4\over 3} (3j+4) (3j+2) (3j+1) (J^-_0 \otimes J^+_0)  \cr
& \qquad 
- 9 (j+1) (j+6) (J^-_0 J^-_0 \otimes J^+_0 J^+_0) 
+ 9 (J^-_0 J^-_0 J^-_0 \otimes J^+_0 J^+_0 J^+_0) \Bigr]
\left| -\onth \right\rangle \otimes |j-1\rangle \,.}}
Similarly, the representation $\H_E$ has a null vector at grade two
that does not involve $J^+_{-1}$; it is given by 
\eqn\appfour{\eqalign{
\N_E = & \Bigl( 
  3 (3j+1) (3j+2) (3j+4) J^3_{-1} J^3_{-1} 
+ (3j+1) (3j+2) (3j+4) (6j-1) J^3_{-2} \cr
& \qquad 
+ 81 j J^-_{-1} J^-_{-1} J^+_0 J^+_0 
+ 3 (3j+4) (9 j^2 - 3 j - 8 ) J^-_{-2} J^+_0 \cr
& \qquad
+ 18 (3j+1) (3j+4) J^-_{-1} J^3_{-1} J^+_0 
+ 3 (3j+1) (3j+2) (3j+4) J^+_{-2} J^-_0 \Bigr) \left| j \right\rangle 
\,.}}
Considering $0=|-{4\over 3}\rangle \otimes \N_E$, and using the 
algorithm \simple\ we then obtain (again after some algebra)
the relation 
\eqn\appfive{\eqalign{
0 = & \Bigl[ 
- {2\over 9} (3j+1) (3j+2) (3j+4) (3j-1) (3j-2)
          (\bbbone\otimes\bbbone) \cr
& \qquad 
+ 4 (3j+1) (3j+2) (3j+4) (2j+1) (J^-_0 \otimes J^+_0)  \cr
& \qquad 
- 9 (3j+4) (3j^2+13j+2) (J^-_0 J^-_0 \otimes J^+_0 J^+_0) \cr
& \qquad
+ 81 j (J^-_0 J^-_0 J^-_0 \otimes J^+_0 J^+_0 J^+_0)
\Bigr] \left| -\onth \right\rangle \otimes |j-1\rangle \,.}}
Remarkably, the two equations \appthree\ and \appfive\ are linearly
independent except for $j=1$. For $j\ne 1$, we can therefore combine
these two equations to obtain 
\eqn\appsix{\eqalign{
0 = & \Bigl[
{8\over 9} (3j+1) (3j-1) (3j-2)  (\bbbone\otimes\bbbone) 
- 4 (3j+1) (3j+4) (J^-_0 \otimes J^+_0) \cr
& \qquad 
+ 36 (J^-_0 J^-_0 \otimes J^+_0 J^+_0)
\Bigr] \left| -\onth \right\rangle \otimes |j-1\rangle \,.}}
On the two-dimensional space spanned by 
$|-\onth\rangle\otimes |j-1\rangle$ and 
$(J^-_0\otimes J^+_0) |-\onth\rangle\otimes |j-1\rangle$ we have
determined the action of $L_0$ using the comultiplication formula 
\eqn\comullzero{
\Delta(L_0)=\left(L_{-1}\otimes \bbbone\right) 
            +\left( L_0\otimes\bbbone \right)
            + \left(\bbbone\otimes L_0 \right)}
and the Sugawara expression for $L_{-1}$. Using \appsix\ to rewrite 
$(J^-_0 J^-_0\otimes J^+_0J^+_0 ) |-\onth\rangle\otimes |j-1\rangle$
in terms of the above two states, we have found that the action of 
$L_0$ is described by the matrix
\eqn\Lzerom{
L_0 = \pmatrix{ -1 & - {2\over 27} (3j-1) (3j-2) \cr
{1\over (j-\twth)} & (j - \twth)} \,.}
This matrix has eigenvalues $-\onth$ and $j-{4\over 3}$, as claimed.  

On the other hand, for $J^3_0=-\onth$, the relevant space is
three-dimensional, and can be taken to be spanned by 
$|-\onth\rangle\otimes |0\rangle$, 
$|-{4\over 3}\rangle\otimes |1\rangle$ and 
$|-{7\over 3}\rangle\otimes |2\rangle$. Using \appthree\ (or
\appfive) we then find that the action of $L_0$ is described by the
matrix 
\eqn\Lzeromp{
L_0 = \left(
\eqalign{- {5\over 3} & \quad - {4 \over 27} \qquad 0 \cr
{27 \over 2} & \qquad \,\, {1\over 3} \quad - {100\over 27} \cr
- {27 \over 8} & \qquad \,\,\, 0 \qquad\,\, {4 \over 3}} \right)\,.}
This matrix is then conjugate to the Jordan normal form matrix
\eqn\Lzerompp{
L_0 = \pmatrix{
\twth & 0 & 0 \cr
0 & -\onth & 1 \cr
0 & 0 & -\onth}\,.}
The two states that span the Jordan block with $h=-\onth$ are
explicitly given by 
\eqn\appseven{\eqalign{
\left(-\onth,-\onth\right)_1 & = 
{11 \over 9} \left(\left| -\onth\right\rangle \otimes |0\rangle\right) 
- { 7\over 2} \left(\left| -{4\over 3}\right\rangle \otimes |1\rangle
\right) 
+ {9\over 8} \left(\left| -{7\over 3}\right\rangle \otimes
|2\rangle\right)   \cr
\left(-\onth,-\onth\right)_2 & = 
-{10 \over 9} \left(\left| -\onth\right\rangle \otimes |0\rangle\right) 
+ 10 \left(\left| -{4\over 3}\right\rangle \otimes |1\rangle\right) 
- {9\over 4}\left( \left| -{7\over 3}\right\rangle \otimes
|2\rangle\right) \,.}} 
When we quotient further to obtain the `highest weight space' we have
the additional relations
\eqn\appeight{\eqalign{
0 & = - \left(\left| -\onth\right\rangle \otimes |0\rangle\right) + 
{9\over 2} \left(\left| -{4\over 3}\right\rangle \otimes
|1\rangle\right)  \cr
0 & = {2\over 9} \left(\left| -\onth\right\rangle \otimes
|0\rangle\right) - 
3 \left(\left| -{4\over 3}\right\rangle \otimes |1\rangle\right) 
+ {9\over 10}  \left(\left| -{7\over 3}\right\rangle \otimes
|2\rangle\right)  \,.}}
Using the relations in \appeight\ it is then easy to see that 
$\left(-\onth,-\onth\right)_2 \cong 0 $ in the highest weight space.

\appendix{B}{The highest weight space of $(D^+_{-\onth}\otimes\R_0)$}

In this appendix we want to determine the highest weight space of the
fusion product in \asstwo. Using \simple\ it is easy to see that this
highest weight space can be taken to be contained in 
$(D^+_{-\onth})^{(0)}\otimes \R_0$. Naively, one may further think 
that one can also restrict $\R_0$ to its highest weight space;
however, this is not quite correct since the $L_0$ spectrum of $\R_0$
is unbounded from below. However, we can derive the recursion
relations  
\eqn\assend{\eqalign{
\left(J^+_0 |j\rangle \otimes (m,m) \right) & \propto
 \left( |j\rangle \otimes (m+1,m+1) \right) \cr
\left(J^+_0 |j\rangle \otimes \omega \right) & \propto
\left( |j\rangle \otimes (1,1) \right) \cr
\left(J^-_0 |j\rangle \otimes (-m,m) \right) & \propto
\left( |j\rangle \otimes (-m-1,m+1) \right)\cr
\left(J^-_0 |j\rangle \otimes \omega \right) & \propto
\left( |j\rangle \otimes (-1,1) \right) \,,}}
where $m\ne 0$, $|j\rangle\in (D^+_{-\onth})^{(0)}$, and we have used
the same notation as in section~7 as well as $\Omega=(0,0)$; none of
the proportionality constants vanishes. If $m\geq 1$, these relations
imply that  
\eqn\assendtwo{\eqalign{
\left(|j\rangle \otimes (m,m) \right) & \propto 
\left( |j+m\rangle \otimes \omega \right) \cr
\left(|j\rangle \otimes (-m,m) \right) & \propto 
\left( |j-m\rangle \otimes \omega \right)\,,}}
where the proportionality constants are non-zero, and $|l\rangle=0$ if 
$l>-\onth$. These states have the same $J^3_0$ spectrum as
$(D^+_{-\onth})^{(0)}$. 

It remains to analyse the states with $m\leq 0$. First of all, it
follows from \assend\ together with the fact that $|-\onth\rangle$
is annihilated by $J^+_0$ that 
\eqn\assendthree{
\left(|j\rangle \otimes (m,m) \right) \propto  0 \qquad \hbox{for
$m\leq 0$}\,.}
In particular, this implies that $|j\rangle\otimes \Omega\cong 0$ for
all $j$. It then also follows from \assend\ that
\eqn\assendfour{
|j\rangle \otimes (-m,m) \cong 0 \qquad \hbox{if $j-m\leq -\onth$}}
since we can use \assend\ repeatedly to rewrite 
$|j\rangle \otimes (-m,m)$ in terms of 
$|j-m\rangle\otimes\Omega\cong 0$. However, if $j-m\geq \twth$, this
argument breaks down and we can only conclude that 
\eqn\assendfive{
\left(|j\rangle \otimes (-m,m)\right) \propto 
\left(\left|-\onth\right\rangle \otimes 
\left(j-m+\onth,-j+m-\onth\right) \right)\,.}
Thus we get one additional state for each $J^3_0$ eigenvalue of
$\twth+n$ where $n\geq 0$; these combine with the states in
\assendtwo\ to give the complete highest weight space of
$\R_{-\onth}$. 
\vskip4pt

\listrefs

\bye